# Radiation Exposures and Compensation of Victims of French Atmospheric Nuclear Tests in Polynesia

Sébastien Philippe[1,*] Sonya Schoenberger[2,3], Nabil Ahmed[4]

[1] Program on Science and Global Security, Princeton University, 221 Nassau St, 2nd floor, Princeton NJ 08540, USA
[2] Department of History, Stanford University, 450 Serra Mall Bldg 200, Stanford, CA 94305, USA
[3] Yale Law School, 127 Wall St, New Haven, CT 06511, USA
[4] Faculty of Architecture and Design, Norwegian University of Science and Technology, Alfred Getz vei 3, 7491 Trondheim, Norway

*To whom correspondence should be addressed: sebastien@princeton.edu

**Abstract**

Between 1966 and 1974, France conducted 41 atmospheric nuclear weapon tests in Polynesia. Radioactive fallout impacted downwind atolls and islands leading to the external and internal exposure of the local populations to ionizing radiation. By law, individuals who were exposed to radiation in the context of these tests can file compensation claims with the French government if they develop certain radiogenic cancers. A claimant who meets the basic eligibility criteria (place, time, and type of disease) automatically benefits from the presumption of a causal link between radiation exposure and the development of their illness, unless the exposure is deemed too low. Since 2017, an effective dose threshold of 1 mSv per year has been used in the claims adjudication process. Decisions as to whether claimants have met this 1 mSv threshold in a given year are often made on the basis of data from government dose reconstruction studies carried out in 2005 and 2006. Using new information available from recently declassified documents on the radiological impact of French atmospheric nuclear tests in Polynesia, as well as atmospheric transport modeling of radioactive fallout, this article shows that maximum doses to the public for key atmospheric tests may have been underestimated by factors of 2 to 10 and estimates that the total population exposed above the compensation threshold of 1 mSv/yr could be greater than ~110,000. This latter result corresponds to about 90% of the entire population of French Polynesia at the time of the atmospheric tests. Integrating these updated dose estimates into the claim adjudication process would enlarge the pool of eligible claimants by a factor of 10. Also discussed are the legal and policy implications of these findings.

# INTRODUCTION

Between 1966 and 1974, France conducted 41 nuclear tests on the Moruroa and Fangataufa atolls in French Polynesia. A combined yield of 10.1 Mt of TNT equivalent was detonated at or above ground level between July of 1966 and September of 1974 (see supplementary table S1). In several cases, tests produced significant radioactive fallout on communities downwind, including on the Gambier archipelago, the atoll of Tureia, and Tahiti, the most populated island of Polynesia.[1,2] Exposure to radioactive fallout from nuclear tests can lead to adverse health effects, including increased risks of developing radiation-induced cancers.[3,4,5] Compensation regimes are one way for victims of atomic testing to seek redress for health problems that may be related to their exposure to such fallout.[6]

In January 2010, the French government created a single unified compensation mechanism for victims under the "Law regarding the recognition and compensation of victims of French nuclear testing," colloquially known as the "*Loi Morin*."[7] The Law established a committee, the *Comité d'Indemnisation des Victimes des Essais Nucléaires* (CIVEN), for processing claims from veterans, former workers at the test sites, and members of the public.[8] The *Loi Morin* stipulates that individuals suffering from a designated list of 23 (originally 21) potentially radiation-induced cancers who lived or sojourned in specific geographic areas where France conducted nuclear tests in Algeria and French Polynesia during the period of testing should benefit from a presumption of causality between the tests and their illness. The original version of the law stated that this presumption of causality should be upheld *unless* CIVEN considers the probability of such a causal link to be "negligible" due to the nature of their disease and the level of exposure.[9]

---

[1] Ministère de la Défense, « La dimension radiologique des essais nucléaires français en Polynésie : à l'épreuve des faits », 2006.

[2] Gérard Martin (dir.), Les atolls de Mururoa et Fangataufa. Les expérimentations nucléaires, aspect radiologique, Saclay, Commissariat à l'énergie atomique, 2007.

[3] National Research Council. (2003). Exposure of the American population to radioactive fallout from nuclear weapons tests: a review of the CDC-NCI draft report on a feasibility study of the health consequences to the American population from nuclear weapons tests conducted by the United States and other nations. (National Academies Press, Washington DC).

[4] Simon, Steven L., and André Bouville. "Health effects of nuclear weapons testing." *The Lancet* 386, no. 9992 (2015): 407-409.

[5] Matanoski, Genevieve M., John D. Boice Jr., Stephen L. Brown, Ethel S. Gilbert, Jerome S. Puskin, and Tara O, Toole. "Radiation Exposure and Cancer: Case Study." American Journal of Epidemiology. Vol 154, No. 12, Supplement, 2001.

[6] Congressional Research Service. "The Radiation Exposure Compensation Act (RECA): Compensation related to Exposure to Radiation from Atomic Weapons Testing and Uranium Mining", 2021.

[7] Loi n° 2010-2 du 5 janvier 2010 relative à la reconnaissance et à l'indemnisation des victimes des essais nucléaires français, https://www.legifrance.gouv.fr/eli/loi/2010/1/5/DEFX0906865L/jo/texte.

[8] Loi n° 2013-1168 du 18 décembre 2013 relative à la programmation militaire pour les années 2014 à 2019 et portant diverses dispositions concernant la défense et la sécurité nationale, JORF n°0294 du 19 décembre 2013, Article 54.

[9] Loi n° 2010-2 du 5 janvier 2010 relative à la reconnaissance et à l'indemnisation des victimes des essais nucléaires français, JORF n°0004 du 6 janvier 2010.

Between 2010 and 2017, the French Ministry of Defense and CIVEN received a total of 1039 applications and awarded compensation in only 31 cases – an overall rejection rate of 97%.[10] In light of these statistics, the French legislature amended the *Loi Morin* through the "Loi EROM" to eliminate the "negligible risk" exception in February 2017.[11] Shortly thereafter, in June 2017, the French *Conseil d'Etat,* France's highest court for matters of public administration, issued an opinion stating that the presumption of causality could only be overturned if the pathology in question resulted *exclusively* from a cause other than ionizing radiation, or if the claimant was not exposed to *any* amount of ionizing radiation.[12] This decision effectively mandated the compensation of all applicants meeting the *Loi Morin*'s basic eligibility criteria—that is, all persons suffering from one of the enumerated cancers who were present in French Polynesia during the period of nuclear testing.

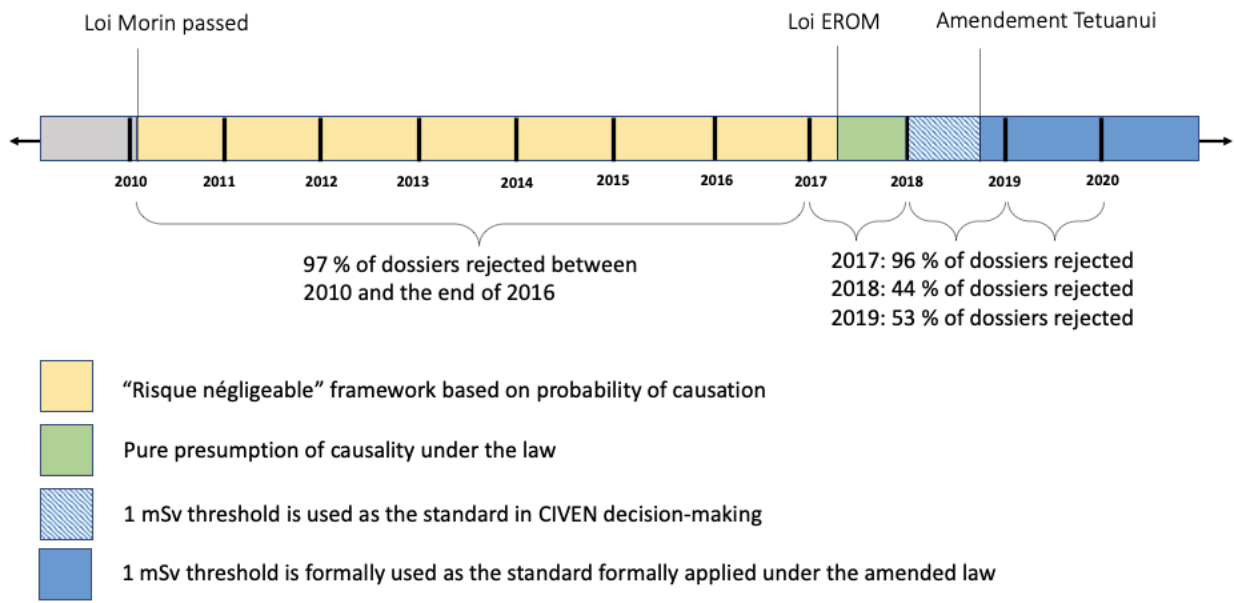

**Figure 1. Timeline of the different methodologies in force for the compensation of victims from French nuclear tests and corresponding claim rejection rates.** *Data from 2019 CIVEN Activity Report.*

The 2017 *Loi EROM* tasked a parliamentary commission with developing a new methodology designed to limit compensation to "only those cancer cases which were caused by French nuclear tests."[13] Based on the commission's recommendations, CIVEN replaced the original "negligible risk" exception with an annual effective dose threshold requirement of 1 mSv in one year – the maximum annual level of ionizing radiation exposure set for the general public under the French Public Health Code.[14] Notably, this 1 mSv/yr threshold drew upon the recommendations of the

---

[10] CIVEN, Rapport d'activité 2019, https://www.gouvernement.fr/sites/default/files/contenu/piece-jointe/2020/07/rapport_dactivite_2019_avec_annexes.pdf, pp. 11-12.
[11] Loi n° 2017-256 du 28 février 2017 de programmation relative à l'égalité réelle outre-mer et portant autres dispositions en matière sociale et économique (« Loi EROM »).
[12] Conseil d'Etat Avis n° 409777, Jun. 28, 2017, JORF n°0154 du 2 juillet 2017.
[13] Loi n° 2017-256 du 28 février 2017 de programmation relative à l'égalité réelle outre-mer et portant autres dispositions en matière sociale et économique (« Loi EROM »), Article 113.
[14] Loi n° 2018-1317 du 28 décembre 2018 de finances pour 2019, JORF n°0302 du 30 décembre 2018, Article 232(2°)(b), referencing Article L. 1333-2(3°) of the French Public Health Code. *See also* Public Health Code Article

International Commission on Radiological Protection (ICRP), which sets out acceptable bands of radiation exposure based on the level of individual and societal benefit associated with such exposures. The ICRP recommends a ceiling of 1 mSv/year in cases in which exposed individuals do not directly benefit from exposure to radiation, but in which such exposures "may be of benefit to society"[15] – a difficult case to make for exposure to nuclear weapon fallout. In June 2020, the French legislature extended this 1 mSv/year threshold retroactively, through an amendment to the COVID-19 health care crisis law, to govern the adjudication of all compensation applications.[16] (See figure 1).

Today, to assess whether a claimant meets the new effective dose threshold of 1 mSv/yr under the conditions of the *Loi Morin,* CIVEN relies upon tables of reconstructed effective and thyroid dose estimates produced by the Military Applications Division of the French *Commissariat à l'Energie Atomique* (CEA/DAM) in 2006.[17] These dose estimates are central to the compensation process and essentially dictate which segments of the Polynesian population present at the time of the nuclear tests are able to receive financial compensation from the French government.

In its most recent paper on its compensation mechanism, CIVEN claims that both the 2006 CEA/DAM data and the methodology through which they were obtained were "validated" by international experts commissioned by the International Atomic Energy Agency (IAEA).[18] However, the 2009 IAEA report referred to by CIVEN made it clear that its review could not and did not assess the validity of any data or computations used to produce the retrospective dose estimates. Its authors simply assumed all data and computation were correct and only provided qualitative assessments of the dose reconstruction methodology.[19] Thus, more than ten years after the *Loi Morin* was enacted, an independent assessment of the dose tables used in adjudicating claims has still not been conducted.

Here, we review both the data and computations of the 2006 dose reconstruction study carried out by the CEA/DAM that form the scientific basis of the adjudication process for compensation claims made by members of the public. We do so in light of new information available in hundreds of primary source documents declassified in 2013, which provide historical measurement data relating to the internal and external radiation exposures of local populations during the period of France's atmospheric testing campaign in French Polynesia (1966-1974).

---

R1333-11 (which states that 1mSv, with the exception of certain particular cases, the annual effective dose limit for the general population resulting from nuclear activities).

[15] ICRP, "The 2007 Recommendations of the International Commission on Radiological Protection," *Annals of the ICRP,* Vol. 37 (2007), p. 96, para. 239.

[16] Loi n° 2020-734 du 17 juin 2020 relative à diverses dispositions liées à la crise sanitaire, à d'autres mesures urgentes ainsi qu'au retrait du Royaume-Uni de l'Union européenne (effectively abrogating the January 27, 2020 Conseil d'Etat Avis n° 432578).

[17] Ministère de la Défense, « La dimension radiologique des essais nucléaires français en Polynésie : à l'épreuve des faits », 2006.

[18] CIVEN, « La Méthodologie suivie par le CIVEN », Annexe à la délibération N°2020-1 du 22 juin 2020, https://www.gouvernement.fr/sites/default/files/contenu/piece-jointe/2020/06/methodologie_suivie_par_le_civen_-_22_juin_2020.pdf, p. 12.

[19] L'Agence Internationale de l'Énergie Atomique (IAEA), « Rapport sur l'examen par des experts internationaux de l'exposition du public aux radiations en Polynésie française suite aux essais atmosphériques nucléaires français, » (Septembre 2009 – Juillet 2010), p. 23.

We identify important shortcomings with the data used as the basis for the past and ongoing adjudication of compensation claims. In addition, we reconstruct the spatial extent of key nuclear fallout events using atmospheric transport modeling for radioactive particles. This allows us to evaluate the impact of the tests on islands and remote atolls in Polynesia for which no or limited historical radiological surveillance data is available.

Taken together, our results suggest that, while the figures currently used by the French government indicate ~11,000 residents of French Polynesia meet the 1 mSv/yr threshold, the true pool of individuals potentially eligible for compensation could be up to ten times greater. This result is primarily obtained from a reconstruction of the radioactive cloud from the 1974 "Centaure" test, which directly impacted the most populated areas of Polynesia including Tahiti, Moorea and the Leeward islands.

In what follows, we provide a brief history and summary of publicly available official dose estimates from French nuclear tests in Polynesia, present our analysis for three specific atmospheric nuclear tests, and discuss the legal and policy implications of our findings.

**Publicly available estimates of effective doses to the population**

The first official estimates of population exposures from the 1966-1974 French atmospheric nuclear tests were published in a 1997 paper as part of a compendium of documents provided by the French government to the International Atomic Energy Agency (IAEA).[20] After stressing that the decision to conduct a nuclear weapons test was conditioned on being able to demonstrate "the insignificance of the health impact on inhabited islands," that "most fallout concerned uninhabited areas," and that "the doses induced by each test are negligible for the vast majority of the tests," author G. Bourges identified five test events for which consequences on inhabited locations were the most significant.[21] These were: the Aldebaran (7/2/1966) and Phoebe (8/8/1971) tests on the Gambier Islands; the Acturus (7/2/1967) and Encelade (6/12/1971) tests on the Tureia atoll; and the Centaure (7/17/1974) test on Tahiti, the most populated island of French Polynesia (See Table S1 for a list of all French atmospheric tests in the Pacific). For each test, effective doses (stochastic health risk to the whole body) from external exposure (sum of groundshine and cloudshine), inhalation, and ingestion were computed from historical measurements (See Table S2). The measurements included activity in the air, on the ground, dose rates, as well as contamination of water, milk and other foodstuffs, that were presented as "annual maximum exposure." No detail about the calculations or references to source documents from which measurement data were derived was shared by the author at the time.

In 2006, the French Ministry of Defense published a 477-page report titled "The Radiological Dimension of French Nuclear Tests in Polynesia: Facing the Facts."[22] The document provides a relatively detailed account of the French atmospheric nuclear test program, the resulting environmental contamination, information on fallout as well as data and information on the organization of radioprotection at the test sites. The report is predicated on "the analysis of existing documents to extract the most representative information on the subject from 1966 to this day [2006]," and includes the results of new dose reconstruction studies from key atmospheric tests on the populations of the Gambier, Tureia, and Tahiti islands. The doses (reproduced in Table 1) include those published in 1997 (with a few additional tests), and new effective and thyroid dose estimates produced by the Military Application Directorate of the French Atomic Energy Commission (CEA/DAM) in 2005 and 2006. Unfortunately, the report does not include or cite primary source material or the detail of dose computations, complicating the task of reviewing the dose estimates presented.

---

[20] Bourges G. Study of the radiological situation at the Atolls of Mururoa and Fangataufa. Paris: CEA/DAM/DRIF/DASE; 1997. Radiological consequences of the atmospheric tests on the islands of French Polynesia from 1966 to 1974.
[21] Bourges G. Study of the radiological situation at the Atolls of Mururoa and Fangataufa, p. 933.
[22] French Ministry of Defense, 2006. The Radiological Dimension of French Nuclear Testing in Polynesia: facing the facts. Ministry of Defense, Paris (in French).

**Table 1.** CEA 2006 reconstruction of effective whole-body and thyroid dose ranges received by members of the public from French atmospheric tests in Polynesia. Doses reconstructed by the CEA in 2006 are the basis for current compensation decisions. They include "minimum" and "maximum" bounds with the exception of the Centaure test. (Source: MinDef, 2006, p. 293).

| Place | Year | Test | CEA 2006 Dose Reconstruction (mSv) | | | |
|---|---|---|---|---|---|---|
| | | | **Effective dose child age 1-2** | **Thyroid child age 1-2** | **Effective dose Adult** | **Thyroid Adult** |
| GAMBIERS | 1966 | ALDEBARAN | 3–10 | 4–78 | 3–7 | 2–40 |
| | 1966 | RIGEL | 0.4–0.71 | 4.6–7.8 | 0.1–0.23 | 1.1–2.1 |
| | 1971 | PHOEBE | 0.5–7.9 | 4.8–98 | 0.2–2.6 | 1.3–26.7 |
| TUREIA | 1966 | RIGEL | 0.1–0.23 | 0.6–2 | 0.06–0.15 | 0.15–1 |
| | 1967 | ARCTURUS | 0.9–4 | 2–38 | 0.79–3.2 | 0.9–25 |
| | 1971 | ENCELADE | 1.5–3.5 | 4–27 | 1.3–1.9 | 1–8 |
| TAHITI (key districts) | | | | | | |
| Pirae (Papeete) | 1974 | CENTAURE | 1.2 | 14 | 0.5 | 4 |
| Hitiaa | 1974 | CENTAURE | 5.3 | 49 | 2.6 | 12 |
| Teahupoo and Taravao | 1974 | CENTAURE | 4.5 | 40 | 3.6 | 16 |

In 2009, the French Government asked the IAEA to commission an independent review to validate the 2006 retrospective dose calculations.[23] However, the review's depth was hampered by its limited scope. At the outset the IAEA report explained that their objective "was not to conduct a new detailed study or a new computation of doses from the data presented [by the French government]." Moreover, its conclusions are based "on the idea that all information, all calculations, and all data provided in the 2006 report and other supplementary reports are correct."[24]

Supplementary CEA reports describe the hypotheses and calculations behind the data summarized in Table 1. These documents, which were not included in the more comprehensive 2006 Ministry of Defense report, contain the details of the dose-computation methodology,

---

[23] IAEA, Rapport sur l'examen par des experts internationaux de l'exposition du public aux radiations en Polynésie française suite aux essais atmosphériques nucléaires français, September 2009 – July 2010, pp 1- 214.
[24] Ibid., p. 23

source term estimation, and data points such as air activity, ground deposition, and dose rate as well as water, milk and foodstuff contamination levels used in the calculations.[25,26,27,28,29,30,31]

Without access to original historical data and documents, the IAEA-commissioned experts concluded that actual population exposures could have been even lower than the official French government estimates.[32] In particular, the reviewers believed dose estimates were based on "the highest measurement results and relatively conservative assumptions."[33] Yet, they also noted that uncertainties about dose estimates were particularly important and that it was impossible to quantify them, because of "the absence of direct measurements of specific nuclides in all media, for all tests and all sites."[34] Despite the reviewers' admission that their study had serious limitations, official CIVEN documents currently state that for the period of atmospheric testing, doses for the public are available in "table format in a 2006 CEA study, whose methodology and results were validated by an international working group mandated by the IAEA."[35]

Below, we present three case studies with a focus on dose calculation from key atmospheric nuclear tests and in the process review dose reconstruction reports from the French government, highlighting uncertainties and possible errors that are typically associated with such efforts.[36] As we show, such potential errors and omissions can contribute to the underestimation of population exposures. We base this analysis on new information available in recently declassified government documents, which describe dose evaluations from the 1960s and 70s (see supplementary materials) and including some of the original data that formed the basis of the 2006 CEA retrospective calculations.

---

[25] CEA/DAM, Calcul de l'impact dosimétrique des retombées de l'essai ALDEBARAN sur les îles Gambier, Dossier d'étude technique, May 2, 2006.
[26] CEA/DAM, Calcul de l'impact dosimétrique des retombées de l'essai ENCELADE à Tureia, Dossier d'étude technique, May 2, 2006.
[27] CEA/DAM, Calcul de l'impact dosimétrique des retombées de l'essai CENTAURE à Tahiti, Dossier d'étude technique, May 2, 2006.
[28] CEA/DAM, Calcul de l'impact dosimétrique des retombées de l'essai PHOEBE sur les îles Gambier, Dossier d'étude technique, July 13, 2006.
[29] CEA/DAM, Calcul de l'impact dosimétrique des retombées de l'essai ACTURUS à Tureia, Dossier d'étude technique, August 29, 2006.
[30] CEA/DAM, Calcul de l'impact dosimétrique des retombées de l'essai RIGEL sur les îles Gambier, Dossier d'étude technique, September 20, 2006.
[31] CEA/DAM, Calcul de l'impact dosimétrique des retombées de l'essai RIGEL à Tureia, Dossier d'étude technique, September 20, 2006.
[32] IAEA, Rapport sur l'examen par des experts internationaux de l'exposition du public aux radiations en Polynésie française suite aux essais atmosphériques nucléaires français, September 2009 – July 2010, p 3.
[33] IAEA, Rapport sur l'examen par des experts internationaux, p. 44.
[34] IAEA, Rapport sur l'examen par des experts internationaux, p. 45.
[35] CIVEN, « La Méthodologie suivie par le CIVEN », p. 12.
[36] Hoffman, F. Owen. Environmental dose reconstruction: Approaches to an inexact science. No. CONF-911282--1. Oak Ridge National Lab., 1991.

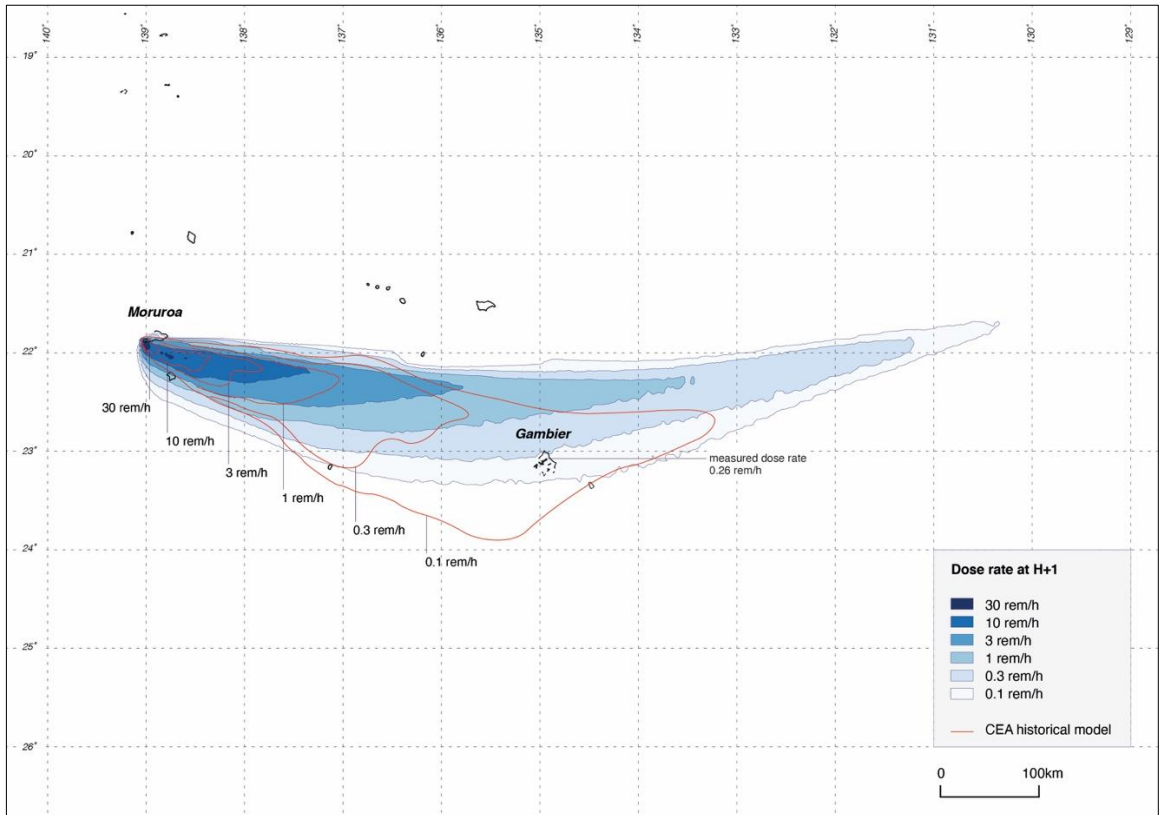

**Fig. 2. Reconstruction of the fallout from the ALDEBARAN nuclear test.** Dose rate contours (solid blue) were obtained using Hysplit (see methods) in order to validate the CEA historical simulation results (red lines) normalized to H+1 after the explosion. Our results are in good agreement with the CEA simulation and with the measured dose rate on the Gambier archipelago.

**RESULTS**

**Case Study 1: Aldebaran fallout on the GAMBIERS archipelago (7/2/1966)**

The first nuclear test conducted at the Centre d'Expérimentation du Pacifique (CEP), codenamed ALDEBARAN, took place on July 2$^{nd}$, 1966. The plutonium device was fired at 15:34 (UT) from a barge anchored on the surface of the Moruroa lagoon (altitude ~10m, water depth=30–40m) and generated a ~28 kT fission yield. The radioactive debris cloud reached an altitude of 9000 m and was pushed away by west-north-west winds.[37] Meteorological data reproduced in a 1967 declassified document indicates the wind directions were measured 3 hours before the test.[38] Gamma detection buoys meant to measure the primary axis of the fallout were deployed in an arc whose sector covered the Gambier islands.[39]

---

[37] Gérard Martin (Coord.), Commissariat à l'Énergie Atomique, "Les Atolls de Mururoa et de Fangataufa, Les Expérimentations Nucléaires : Aspects radiologiques," Rapport CEA - R - 6136 (2007), 164.
[38] SMSR, "Rapport sur l'évolution de la radioactivité en Polynésie due aux retombées des explosions Françaises au Pacifique," Report 8/SMSR/PEL/CD, March 17, 1967, p. 19
[39] Gérard Martin (Coord.), Commissariat à l'Énergie Atomique, "Les Atolls de Mururoa et de Fangataufa, Les Expérimentations Nucléaires : Aspects radiologiques," Rapport CEA - R - 6136 (2007), 165.

The fallout reached the Gambier islands 10 hours and 45 minutes after the test. Dry deposition of radioactive particles took place for about one hour and 20 minutes, leading to a ground deposition of 6.2 $10^7$ Bq/m$^2$ (single measurement of 1.67 mCi/m$^2$).[40] The dose rate at the end of the fallout reached 0.26 mGy/h [26 mrad/h].[41] It rained on the island shortly after, leading to the contamination of rainwater collection systems. Despite ample time for issuing warnings, the population was not alerted about the risks of radiation exposures, although a French minister visiting the island was evacuated before the arrival of the radioactive cloud. Authorities dispatched to measure and evaluate foodstuff contamination on the islands suggested that it might be necessary to minimize the true dose estimates to avoid losing the trust of local inhabitants.[42] The 1966 estimates concluded that the population received effective whole-body and thyroid doses of 7.12 mSv [0.712 rem] and 15.4 mSv [1.54 rem].[43] These numbers were obtained using methodologies and models available at the time and were computed for adults only.

In the 2006 dose reconstruction study of the event, the CEA/DAM computed whole-body and thyroid doses from groundshine, cloudshine, inhalation, and ingestion of radionuclides for multiple age groups.[44] The study developed minimum and maximum estimates of the whole-body and thyroid doses to the local population based on different assumptions.

For groundshine, a dose rate of 0.14 mSv/h was computed at H+11 (i.e. 11 hours after detonation) from the ground deposition data and a radionuclide source term produced by CEA/DAM. This dose rate was about half that measured at the time but considered consistent with the measured ground activity. The CEA then calculated the corresponding external dose up to 6 months (and not up to a year). The result was then multiplied by 2/3 to account for time typically spent outside rather than inside during the day, as opposed to a factor of 0.75 used in documents at the time. One such document shows that significant radioactivity was measured on the clothes of a local inhabitant who was sleeping outside,[45] highlighting the problem of using a correction factor in this particular context to produce a maximum dose estimate.

For cloudshine and inhalation, two different values of deposition velocity ($V_d$= $10^{-2}$ and $10^{-1}$ m/s) were used by the CEA to generate a minimum and maximum total integrated activity in air. These values are coherent with the dry deposition velocities of particles in the range of 6 to 20 μm in diameter with density ~2g/cm$^3$.[46] Given that the test occurred on a barge anchored at the surface of the Moruroa lagoon and the relative proximity of the Gambier islands to ground zero, it is likely that a wider range of particle sizes reached the atoll, yet no uncertainties associated with this parameter were taken into account.

---

[40] SMSR, "Bilan des mesures physiques concernant Mangareva, Juillet 1966," Report 14/SMSR/PEL/PAC/S, August 19, 1966, p. 6.
[41] SMSR, "Bilan des mesures physiques concernant Mangareva, Juillet 1966," p. 5.
[42] P. Millon, "Mission de la Coquille aux Gambiers du 2 au 10 juillet 1966," Report MN/CEP/BRO Coquille/SMCB/S, July 10, 1966.
[43] SMCB, "Fiche sur la Synthèse des rapports SMSR/SMCB/SANTE relatifs à la retombée Aldebaran," 1967.
[44] CEA/DAM, Calcul de l'impact dosimétrique des retombées de l'essai ALDEBARAN sur les îles Gambier, Dossier d'étude technique, May 2, 2006.
[45] SMCB, "Rapport préliminaire concernant les résultats obtenus par le BRO "La Coquille" pendant la 1ere demi-campagne," Report 110/CEP/SMCB/S, August 165, 1966.
[46] Gérard Martin (Coord.), Commissariat à l'Énergie Atomique, "Les Atolls de Mururoa et de Fangataufa," 110.

For exposure from the ingestion of contaminated water, the CEA assumed that either no contaminated water was consumed by the inhabitants in the case of the minimum exposure dose estimate or that contaminated water from the main village stream water collection system (814 Bq per liter measured on July 8, corresponding to 14000 Bq/l at H+11) was consumed to compute the maximum dose. While the report of the radiological survey team dispatched to the island described some French government employees drinking bottled water,[47] the local population, who were not made aware of the fallout, had no access to uncontaminated sources of drinking water. Furthermore, isolated households typically relied on rainwater collected in barrels and cisterns from their roofs as their primary source of drinking water in 1966 and as late as 1977.[48,49]

By taking into account rainwater consumption and following the CEA methodology used in other reports where this contamination pathway is computed,[50] we find that the maximum effective and thyroid dose estimates from water consumption for the first month after the fallout could have been underestimated by a factor of 20 (see methods).

Using this set of information, we re-evaluate the 2006 CEA effective and thyroid doses to adults and children on the Gambier Islands. Our results shown in table 2 suggest that maximum dose estimates could have been underestimated by a factor of ~2.5.

---

[47] P. Millon, "Mission de la Coquille aux Gambiers du 2 au 10 juillet 1966," p. 2.
[48] ORSTOM and SMCB, "Rapport d'étude hydrologique aux Iles Gambier," August 1966.
[49] Insee, "Résultats du recensement de la population de la Polynésie française," April 29, 1977.
[50] CEA/DAM, Calcul de l'impact dosimétrique des retombées de l'essai PHOEBE sur les îles Gambier, Dossier d'étude technique, July 13, 2006.

**Table 2.** Effective and thyroid dose estimates for the ALDEBARAN fallout on the Gambier archipelago. Results are based on 2006 CEA data and our re-evaluations (in bold) of the groundshine dose as well as doses from contaminated water consumption. Groundshine and cloudshine contributions to the thyroid, missing from the CEA analysis, were also included.

|  | Effective dose [mSv] | | | | Thyroid dose [mSv] | | | |
|---|---|---|---|---|---|---|---|---|
|  | Child age 1-2 | | Adult | | Child age 1-2 | | Adult | |
|  | Min | Max | Min | Max | Min | Max | Min | Max |
| Groundshine | **3.02** | **8.14** | **3.02** | **8.14** | **2.87** | **7.74** | **2.87** | **7.74** |
| Cloudshine | 0.06 | 0.29 | 0.03 | 0.67 | **0.07** | **0.30** | 0.03 | 0.71 |
| Inhalation | 0.70 | 2.99 | 0.16 | 3.84 | 3.00 | 30.00 | 1.30 | 13.00 |
|  | | | | | | | | |
| Water | **0.48** | **9.60** | **0.12** | **2.40** | **6.00** | **120.00** | **1.30** | **26.00** |
| Vegetables | 0.10 | 1.70 | 0.09 | 1.40 | 1.30 | 19.00 | 1.00 | 14.00 |
| Fish | 0.00 | 0.64 | 0.00 | 0.32 | 0.02 | 7.00 | 0.01 | 3.20 |
| Mollusc | 0.00 | 1.24 | 0.00 | 0.48 | 0.00 | 15.80 | 0.00 | 5.40 |
|  | | | | | | | | |
| Ingestion | 0.58 | 13.18 | 0.21 | 4.60 | 7.32 | 161.80 | 2.31 | 48.60 |
|  | | | | | | | | |
| Internal | 1.29 | 16.17 | 0.37 | 8.44 | 10.32 | 191.80 | 3.61 | 61.60 |
| External | 3.08 | 8.43 | 3.04 | 8.82 | 2.93 | 8.04 | 2.89 | 8.45 |
|  | | | | | | | | |
| **Total [mSv]** | **4.37** | **24.60** | **3.42** | **17.26** | **13.25** | **199.84** | **6.50** | **70.05** |
| CEA 2006 | 3.20 | 9.40 | 3.10 | 6.60 | 4.00 | 78.00 | 2.00 | 37.00 |
| Ratio | 1.36 | 2.62 | 1.10 | 2.61 | 3.31 | 2.56 | 3.25 | 1.89 |

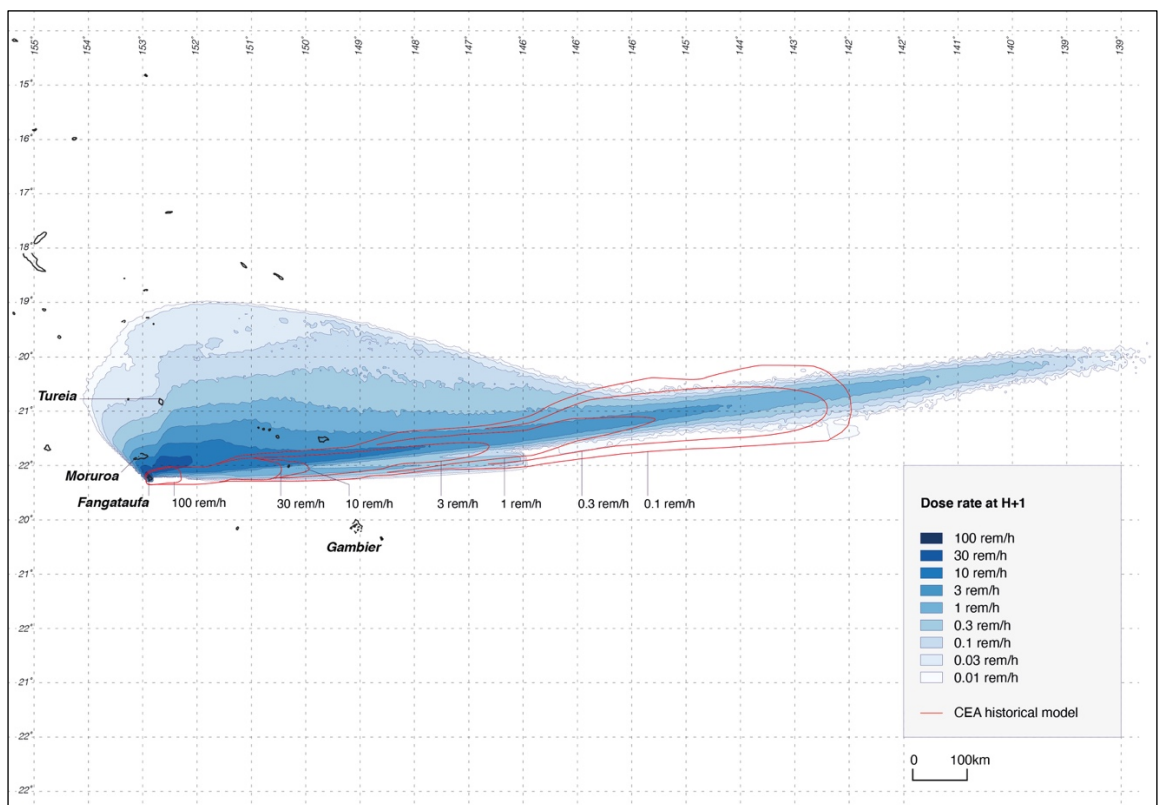

**Fig. 3. Reconstruction of the RIGEL Test Fallout.** Dose rate contours (solid blue) were obtained using Hysplit and normalized to H+1 after the explosion in order to be compared with the CEA historical simulation results (red lines). Unlike the CEA simulation, our results capture the effect of SE lower altitude winds, which pushed the cloud stem over Moruroa and Tureia.

**Case study 2: RIGEL fallout on Tureia and the Gambier Archipelago (9/24/1966)**

The RIGEL test took place on September 24, 1966 from a barge anchored on the lagoon of the Fangataufa atoll (22° 14' 24" S; 138° 43' 22" W). The test involved an experimental thermonuclear device consisting of a plutonium core surrounded by a thick shell of lithium deuteride.[51] While no significant fusion reaction took place, the device generated a 125 kT yield. The test lead to significant contamination at Fangataufa, which required a major clean up by military personnel.[52] After the explosion, the radioactive cloud head travelled eastwards over the uninhabited Acteon islands. Parts of the cloud stem, however, travelled northwards and landed on Mururoa and Tureia, a feature which our simulation reproduced (see figure 3). Both Tureia and the nearby Gambier islands also experienced radioactive rains two days after the test.

The 2006 CEA dose reconstruction for both the Gambier islands and Tureia are primarily based on a single rain activity measurement. From this data point, dose estimates are computed for groundshine, cloudshine, inhalation and ingestion using activity ratios constructed from data available from other tests. For example, the drinking water activity is calculated from the rainwater activity by using ratios of these two values from the Arcturus (1967) and Encelade (1971) tests, which both impacted Tureia and not the Gambier islands. On top of this, three major assumptions have important impact on the final dose estimates from water consumption, especially in the case of Tureia.

---

[51] Pierre Billaud, Venance Journé, "The Real Story Behind the Making of the French Hydrogen Bomb. Chaotic, Unsupported, but Successful," *Nonproliferation Review*, vol. 15, no 2, 2008, p. 353-372.
[52] SMSR, "Compte rendu de decontamination sur l'atoll de Fangataufa," Report 1/SMSR/DR, 1967.

First, at least two different sets of measurements for the original rain contamination levels appear in two different historical documents. One for the report of the Rigel test, and one for the overall 1966 campaign report.[53,54] The dose reconstruction uses the lowest of the two. Using the largest values of activities raise the maximum dose estimate by ~1.52 and ~2.85 for Gambier and Tureia respectively.

Second, assumptions about the ratios of activities between rain contamination and collected rainwater underestimate maximum doses. Tureia had two kinds of rainwater collection systems: a shared communal cistern and individual household cisterns. For the Encelade test, the latter were contaminated up to ~7.4 times more than the communal cistern.[55] The ratios used for the RIGEL reconstruction, however, are based on measurements from two different type of cisterns (communal for Arcturus, and household for Encelade) that have different collection surfaces and capacities, leading to different dilution factors of the contaminated rain. In the case of Arcturus, no data is publicly available for the family cisterns.[56] This does not mean that they were not contaminated with higher levels of fission products. Taking this possibility into account raises the maximum effective and thyroid doses for the local population by at least a factor of ~7.

Third, for Tureia, the population is assumed to drink half the amount of water of other inhabitants in Polynesia (the core assumption is that adults drink one liter of water and two liters of coconut water per day). This assumption also leads to the underestimation of maximum doses by a factor of 2. It is not used, however, in recent dose reconstruction studies.[57,58]

Together, changes in these three assumptions can affect the maximum dose estimates by a factor of ~10 to 20 depending on the age of individuals (see supplementary tables S6 and S7). This result is not surprising given the large uncertainties involved when relying on such limited data. While the reconstruction studies present both "minimum" and "maximum" dose estimates, our analysis and results clearly show that they should not be taken as true confidence intervals nor as reflecting any uncertainty and sensitivity analysis, which is still missing from the government studies available to date.

---

[53] SMSR, "Compte Rendu de l'Opération RIGEL," Report 27/SMSR/PEL/PAC/S, October 19, 1966, p. 7.
[54] SMSR, "Rapport sur l'evolution de la radioactivité en Polynésie due aux retombées des explosions françaises eu Pacifique," Report 8/SMSR/PEL/CD, March 17, 1967, p. 7.
[55] SMCB. "Etude de la dose absorbée en contamination interne par les habitants de Tureia au cours du mois suivant le tir Encelade,"" Report 126/CEP/SMCB, August 10, 1971.
[56] CEA/DAM, Calcul de l'impact dosimétrique des retombées de l'essai ACTURUS à Tureia, Dossier d'étude technique, August 29, 2006.
[57] IRSN, "Evaluation de l'exposition radiologique des populations de Tureia, des Gambier et de Tahiti aux retombées des essais atmosphériques d'armes nucléaires entre 1975 et 1981," Report IRSN/2019-00498, (2019): 1-43.
[58] Drozdovitch, Vladimir et al. "Thyroid Doses to French Polynesians Resulting from Atmospheric Nuclear Weapons Tests: Estimates Based on Radiation Measurements and Population Lifestyle Data." *Health Physics* 120, no. 1 (2021): 34-55.

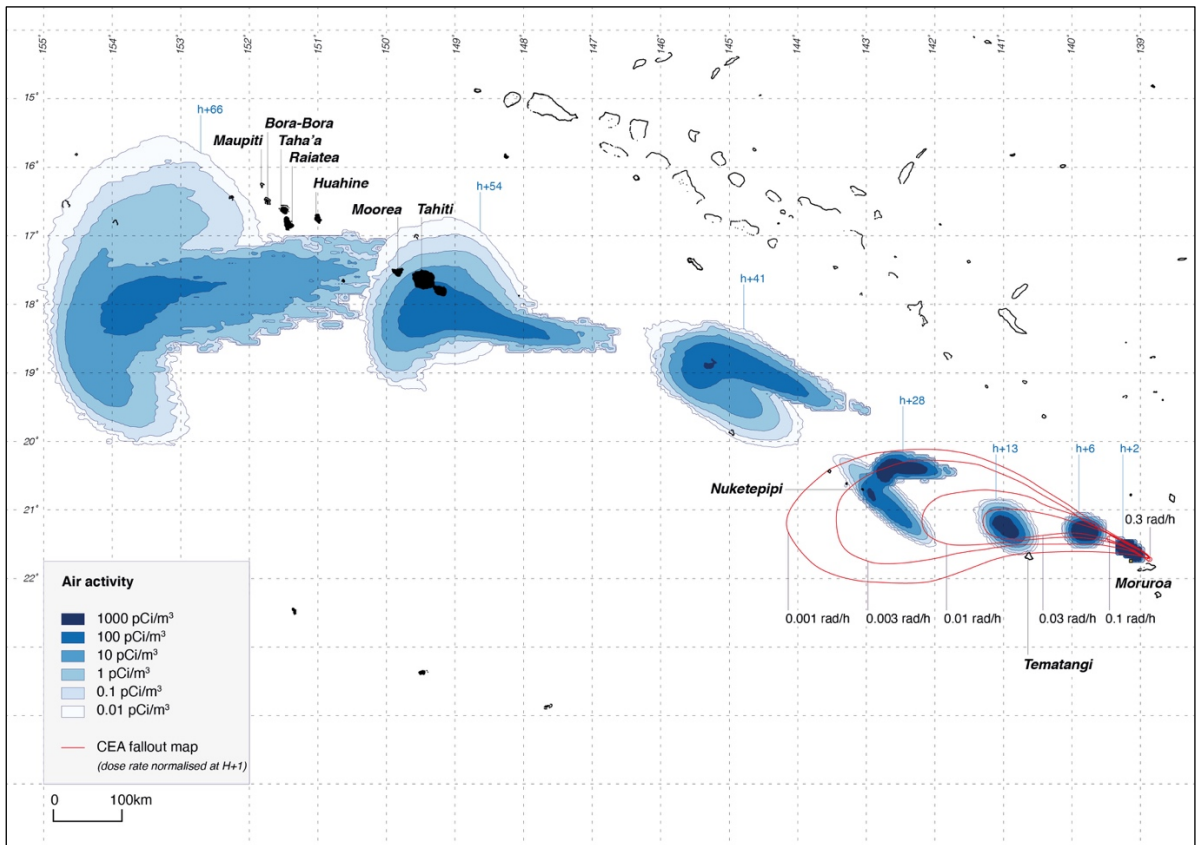

**Fig. 4. Reconstruction of the CENTAURE cloud trajectory over French Polynesia.** Air activities between 0 and 500m were obtained using Hysplit (see methods). We find that the cloud travelled directly from Moruroa to Tahiti in about 2 days before reaching the leeward islands (including Huahine, Raiatea, Taha'a, Bora-bora and Maupiti). Early time steps are consistent with the CEA map of the fallout contour (simulated for the first ~24h). Our simulation underpredicts the measured activity in Mahina, Tahiti by a factor of ~10, however.

**Case study 3: CENTAURE fallout on Society Islands (7/17/1974)**

On July 17, 1974, France detonated a 4-kt plutonium device. The experiment codenamed CENTAURE took place under a balloon 270m above Moruroa (21° 47' 13" S; 138° 53' 32" W).[59] Twelve hours before the test, SMSR (the CEA radiological safety service) predicted the fallout would occur northward from the test site over the atolls of Hao and Tureia, but the potential impact was deemed small enough that the decision to conduct the test was nevertheless approved.[60] After the explosion, French technicians realized the radioactive debris cloud did not reach the expected altitude (the top of the cloud reached 5200m instead of the predicted 8500m). Meanwhile the winds had shifted towards the west.[61] Our reconstruction of the cloud trajectory shows it travelling directly from Moruroa to Tahiti in a straight line (see figure 4 and the video in supplemental materials). The combination of a lower than expected cloud height and heavy rain led to wet and dry deposition on the most inhabited islands of Polynesia.

---

[59] Martin, "Les Atolls de Mururoa et de Fangataufa, Les Expérimentations Nucléaires : Aspects radiologiques," 271.
[60] Ministère de la Défense, "La dimension radiologique des essais nucléaires français en Polynésie," 262.
[61] Martin, "Les Atolls de Mururoa et de Fangataufa, Les Expérimentations Nucléaires : Aspects radiologiques," 746.

The CENTAURE cloud reached the island of Tahiti about two days after the explosion. Despite a good understanding of meteorological conditions and the ability to compute the trajectories of contaminated air masses up to a few days,[62] the population of Tahiti was neither made aware of the arrival of the fallout nor was advised to seek shelter. To the authors' knowledge, no measures were taken to prevent the consumptions of contaminated foodstuffs.

At the Mahina radiological station, the measured dose rate peaked at 390 microrad/h between H+48 and H+54.5. Declassified documents show the total cumulative deposition reached $3.4 \cdot 10^6$ Bq/m$^2$ ($9 \cdot 10^7$ pCi/m$^2$).[63,64] Another recent study (independent from our work) obtained the same cumulative value from the same declassified government documents.[65]

The fallout on Tahiti, which lasted for a couple of days due to wet weather and the particular geography of the island, was relatively heterogeneous with ratios of deposited activity in various parts of the island ranging from 0.1 to 11 times the value measured at the Mahina radiological station (See Fig. 2).

In analyzing declassified government documents, we found multiple errors or omissions in the CEA retrospective dose reconstruction of the CENTAURE test that have important implications for the potential compensation of victims from the most populated area of Polynesia.

First, the CEA used a cumulative deposition of $2.5 \cdot 10^6$ Bq/m$^2$ at Mahina, corresponding to the activity measured on July 19, 1974 only (see Fig. 5.)[66] Because of this lower value, all groundshine, cloudshine and inhalation doses were underestimated by a factor of 3.4/2.5 = 1.36. Second, a careful comparison of the 1974 original map representing the deposited activity on the island of Tahiti[67] and subsequent copies published in official French government publications[68,69,70] (Figure 5) shows that the ground activity for the Papeete/Pirae zone ranged from 0.1 to 0.3 times the value measured at the Mahina station and not 0.1 to 0.2 as the recent government copies suggest. In the dose reconstruction, the CEA used a ratio of 0.13, which is 2.3 times lower than the historical upper bound ratio of 0.3 for this zone as found in the 1974 map.

Combined these two possible errors mean that the groundshine, cloudshine, and inhalation dose for the Pirae/Papeete zone, which housed ~80,000 inhabitants (2/3 of the total Polynesia population) at the time, were underestimated by a factor of ~3.2.

---

[62] SMSR, "Compte rendu de la campagne 1974," Report 11/SMSR/DIR/SD, November 26, 1974, p. 10.
[63] SMSR, « Retombées en Polynésie lors des 1ère et 2éme rafales de la campagne 1974, » Note 2/SMSR/GOEN/SD, August 26, 1974, p. 2
[64] SMSR, Compte rendu campagne 1974, p. 33-38
[65] Drozdovitch, Vladimir, Florent de Vathaire, and André Bouville. "Ground deposition of radionuclides in French Polynesia resulting from atmospheric nuclear weapons tests at Mururoa and Fangataufa atolls." *Journal of environmental radioactivity* 214 (2020): 106176.
[66] CEA, Dossier d'Etude Technique, Calcul de l'impact dosimétrique des retombées de l'essai CENTAURE à Tahiti, 2006, p. 16
[67] SMSR, "Retombées consecutives au tir CENTAURE," Report 101/SMSR.PAC/CD, September 7, 1974, Fig. 7
[68] Bourges, "Study of the radiological situation at the Atolls of Mururoa and Fangataufa," p. 947
[69] Ministère de la Défense, "La dimension radiologique des essais nucléaires français en Polynésie," 217.
[70] Martin, "Les Atolls de Mururoa et de Fangataufa, Les Expérimentations Nucléaires : Aspects radiologiques," 427.

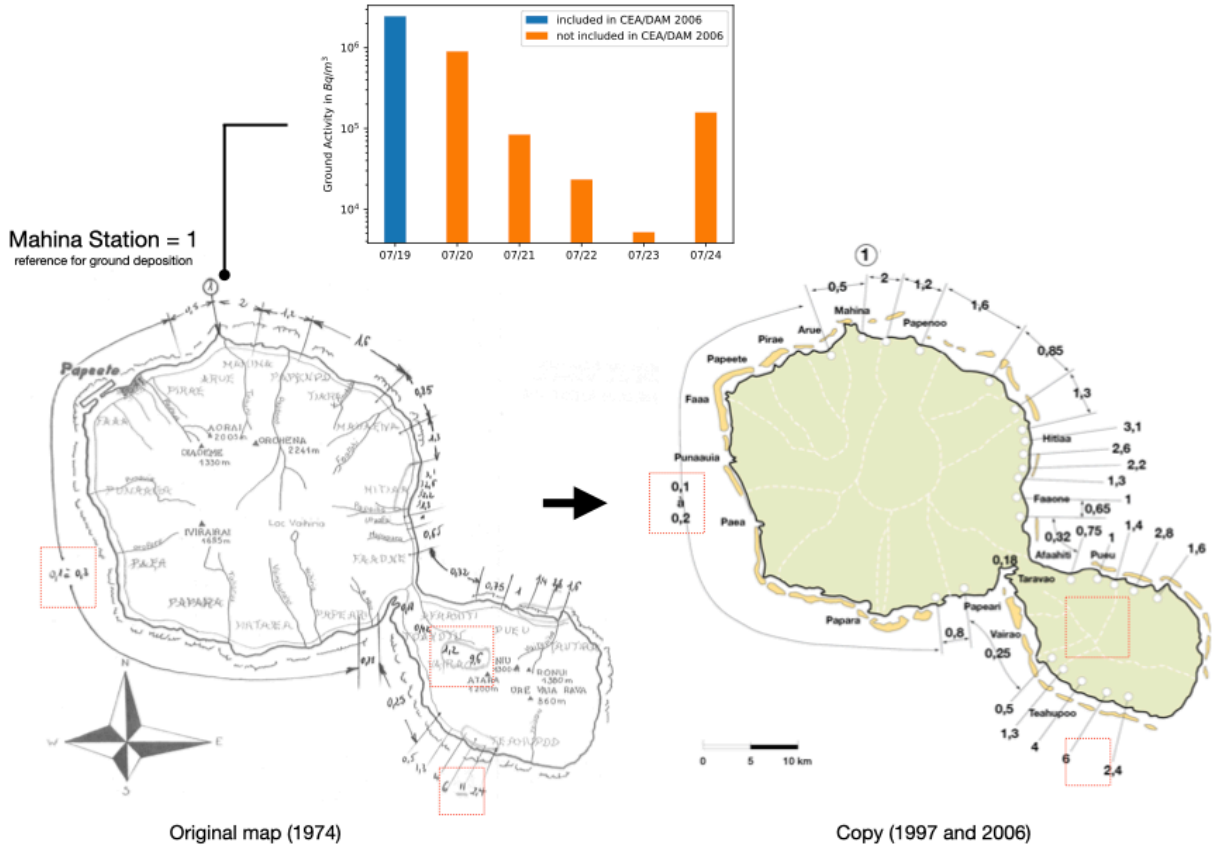

**Fig. 5. Measurements of ground activity in Tahiti following the 1974 Centaure test.** Both maps provide deposition values in relation to the reference data at the Mahina Radiological Control Station. The 2006 official map on the right is the reproduction of the original 1974 map on the left. The highest activity values in Taravao (9.6 times) and Teahupoo (11 times) do not appear on the 2006 version. More importantly, the upper bound for activity in the western part of the main island is modified from 0.3 to 0.2. The CEA 2006 Centaure reconstruction study assume a 0.13 ratio between the Pirae/Papeete zone and Mahina, and as the histogram of the daily deposition measured in Mahina shows, does not take into account ground deposition after July 19.

**Table 3.** Original and revised effective dose estimates in mSv for the Pirae/Papeete area due to the CENTAURE fallout. The 95% upper bound only include measurement uncertainties with regards to the Mahina station ground deposition measurement and nothing else.

|  | **Newborn** | **1-2 yo** | **2-7 yo** | **7-12 yo** | **12-17 yo** | **Adults** |
|---|---|---|---|---|---|---|
| Original effective dose | 0.77 | 1.19 | 0.71 | 0.61 | 0.59 | 0.45 |
| Revised estimates | 0.97 | 2.03 | 1.45 | 1.29 | 1.16 | 0.94 |
| 95% CI upper bound | 1.10 | 2.20 | 1.61 | 1.45 | 1.33 | 1.10 |

With regards to internal contamination, the CEA acknowledges the limited contamination data for foodstuffs, including the absence of radioactivity measurements for vegetables from the Papeete market in July 1974 (CEA 2006, p.22). As a proxy, activities for the Papeete/Pirae region were taken from the Paea market, which had the lowest contamination values on the Island (p. 31). Given the possibility of sourcing vegetables elsewhere, it is equally plausible to assume that vegetables could have originated from nearby Hitiaa (which had the highest measured contamination). Using the latter values raises the maximum effective and thyroid doses from vegetable consumptions by a factor of 2.73 for Papeete. Given the lack of data in declassified government documents, it is difficult to estimate the validity of reconstructed doses from other foodstuffs (meat, fish…). For the purpose of this re-evaluation, we assume they are valid.

Based on our findings, we correct the summary tables for estimated effective and thyroid doses produced by the CEA for the CENTAURE test (see supplementary tables S11 and S12). To produce a confidence interval for our estimates, we assign a 25 percent standard deviation to the Mahina ground deposition measurements. This value is suggested in a CEA publication and apparently typical for this type of measurement.[71] To a first order, the ground deposition measurement error propagates to dose groundshine, cloudshine and inhalation estimates. This allows us to compute a 95% confidence interval for effective and thyroid doses (assuming doses are normally distributed and all other sources of errors to be zero). The revised effective doses results are given for different age groups and presented in Table 3. They suggest that the entire population of Tahiti (~87,500 people at the time) could have received effective (whole-body) doses above 1mSv, the threshold for compensation under the current CIVEN methodology, from the Centaure fallout alone.

Beyond Tahiti, declassified documents show that ground contamination measurements were conducted on other islands including Moorea, Bora-Bora, Raiatea and Huahine[72]. The average ground contamination for these islands were ~0.3 times the value measured at the Mahina reference station. All islands had hotspots reaching up to 0.5 times the reference value, with the exception of Huahine, which showed values up to 2.5 times the reference point. No information about the consumption of contaminated foodstuffs is available for these islands. Given the ground deposition values for Tahiti and known estimates of contamination from foodstuffs, it cannot be ruled out that all the populations on these islands (~17,100 people) also received effective doses greater than 1 mSv.

For other islands where no measurement data is available, our simulation of the Centaure cloud pathway through Polynesia suggests that all the other inhabited Society islands including Maiao, Tahaa, and Maupiti (~4,200 people) were likely impacted by the Centaure fallout.

---

[71] Martin, "Les Atolls de Mururoa et de Fangataufa, Les Expérimentations Nucléaires : Aspects radiologiques," 538.
[72] SMSR, "Retombée consecutive au tir Centaure," ref. 101/SMSR/PAC/CD, September 7, 1974.

**Other dose reconstruction reports and relevant historical estimates**

Beyond the case studies that we have just reviewed, the CEA produced three other reports for the Arcturus (1967), Encelade (1971), and Phoebe (1971) tests. These also present issues with assumptions that are described in the supplementary materials. For these events, we find that maximum effective whole-body and thyroid doses could have been underestimated by factors of 1.5 to 4 and 1.5 to 2.5 respectively (see supplementary materials).

Beyond these cases, to the authors' knowledge, no new government reconstruction studies were conducted in 2005-2006 for the other 35 nuclear tests that took place between 1966 and 1974. This lack of information prevents the possibility to properly address the cumulative effect of multiple exposures within a given year or throughout the entire period of atmospheric testing for the different islands and atolls of Polynesia.

**Summary of results**

Taking available French government data and our new findings into account, as well as population census data from 1967, 1971, and 1977, [73,74] we estimate that the total number of inhabitants who may have received doses greater than 1 mSv/yr to be ~110,000, about 90% of the total Polynesian population in 1974. This estimate includes 87,500 inhabitants in Tahiti, 6,000 in Moorea, 16,000 in the Leeward Islands, and 600 in Gambier and Tureia. This is ten times more people than the current government numbers shown at the beginning of this article in Table 1 and used today by CIVEN suggest.

**DISCUSSION**

Our analysis shows that CIVEN's current methodology has significant shortcomings. The data CIVEN has relied upon to date in adjudicating compensation claims do not take into account accurate upper-bound estimates of claimants' exposure to ionizing radiation during the period of French atmospheric testing. The government studies that serve as the basis for CIVEN's dose estimates do not take measurement and model uncertainties into account and often rely upon poorly-conceived exposure and contamination assumptions. Taken together these methods and practices can be prejudicial to claimants by lowering their chances of possibly receiving compensation.

Case studies 1 and 2 (Aldebaran and Rigel) show how assumptions about the source of drinking water utilized by residents of Tureia and the Gambier archipelago can lead to underestimations of internal exposure to ionizing radiation. Further, Case Study 3 demonstrates that failures to account for the total activity deposited over Tahiti in the wake of the July 1974 Centaure test as well as possible errors in extracting local deposition measurement values from historical maps, and generalized extrapolation of foodstuff contamination levels from measurements taken from less-contaminated parts of the island led to underestimation of the external and internal exposures of inhabitants. Given the high population concentration on the island of Tahiti, this reevaluation of exposures linked to Centaure significantly increases the number of individuals who should be considered eligible for compensation under the current CIVEN methodology.

---

[73] INSEE, "Polynésie Française: Comparaison des résultats des recensements de 1956, 1962 et 1971, et du dénombrement de 1967" September 20, 1971, pp.1-14.
[74] INSEE, "Résultats du recensement de la population de la polynésie française, Annexes" April 29, 1977, table 14.

These findings have significant legal and policy implications. The 2010 *Loi Morin* and its subsequent amendments aim to ensure that individuals whose cancers may have resulted from exposure and contamination linked to French nuclear testing are issued compensation while putting in place certain limitations to avoid over-inclusivity. The Conseil d'Etat, France's highest administrative court, has indicated that in cases of doubt, the law should favor the claimant.[75] Further, the Conseil d'Etat has held that, at the appeals stage, courts dealing with compensation should apply the legal methodology in force that is at best most favorable to a claimant.[76] Thus French courts have set a high bar for overturning the baseline assumption established by the *Loi Morin* that individuals suffering from radiation-induced cancers who meet basic eligibility criteria should be presumptively considered victims of French nuclear testing.[77]

The case studies and analyses based on previously unexamined information gleaned from declassified government documents presented in this paper show the necessity of reevaluating claims previously rejected on the basis of the 1mSv/yr threshold, as claimants, CIVEN, and administrative courts previously relied upon incomplete data and possible dose underestimations, which to our knowledge have never been independently reviewed by any French or international scientific body with the scrutiny they deserve.

In light of our results, both past and future claimants' situations may change when it comes to seeking compensation. Polynesians living in Tahiti in the summer of 1974 whose compensation claims have been rejected by CIVEN on the grounds that they were not exposed to 1 mSv of radiation in a twelve-month period should have the opportunity to seek renewed review of their claims. This should also apply to those born in the winter or spring of 1975 who were *in utero* at the time of the Centaure test.

While those who have already initiated appeals against CIVEN rejections may be able to successfully introduce new exposure estimates in the context of their filings before French administrative courts, claimants who have let the two-month window for appeal in the wake of a rejection from CIVEN expire may find themselves with no clear path to judicial recourse.

---

[75] Conseil d'État Avis n°439003, 06 novembre 2020, https://www.legifrance.gouv.fr/ceta/id/CETATEXT000042506259?tab_selection=all&searchField=ALL&query=439003&page=1&init=true, ¶ 8.

[76] *See, e.g.,* Conseil d'État Avis n°439003 ¶¶ 5-6 (stating that the Bordeaux court of appeals should *not* have applied the more restrictive 1 mSv threshold when a pure presumption of causality under the law of February 28, 2017 could legally be applied instead); ionizing radiation, or if the claimant was not exposed to *any* amount of ionizing radiation); Conseil d'Etat Avis n° 432578, 27 janvier 2020, ¶ 3, https://www.legifrance.gouv.fr/ceta/id/CETATEXT000041485748 (stating that, because the 2018 law made it easier for state to overturn presumption of causality, this limitation applies only to those dossiers submitted *after* this new law came into force).

[77] *See also* Conseil d'Etat Avis n° 409777, 28 juin 2017, <u>JORF n°0154 du 2 juillet 2017</u>, ¶ 4, https://www.legifrance.gouv.fr/jorf/id/JORFTEXT000035072392 (stating that, in the wake of EROM, the presumption of causality could only be overturned if the pathology in question resulted *exclusively* from a cause other than ionizing radiation, or if the claimant was not exposed to *any* amount of ionizing radiation).

Previously unsuccessful claimants could argue that they have just cause for the reconsideration of their claims, given that data to support their claims was either unavailable or obfuscated at the time when their compensation requests were adjudicated. The simplest solution to this problem, however, may well be legislative, and there is precedent here: the February 28, 2017 "Loi EROM," which amended the original *Loi Morin* by eliminating the "risque negligeable" exception to the presumption of causality in favor of claimants, also provided that those whose claims were rejected under the original methodology could resubmit their claims—a limited window of opportunity that expired in December 2020.

From a policy and financial outlays standpoint, the impact of allowing for resubmission of claims would likely not be dramatic: in the first 10 years of the *Loi Morin*, only 285 residents of French Polynesia applied for compensation—and the significant majority of these claims were rejected. The legal implications of this study are likely more numerically significant when considering future first-time claims, in light of our finding that virtually anyone residing on Tahiti in the summer of 1974 may well have been exposed to greater than 1 mSv of ionizing radiation after the Centaure test.

Based on available cancer incidence data from the Polynesian government, we estimate that the current average incidence of cancers that are recognized under the *Loi Morin* to be radiogenic among people born before 1975 in French Polynesia is about 0.4% per year.[78,79] Thus we might expect at most about 350 new cases of cancers per year among the 110,000 people residing in the Society Islands in 1974 (the majority in Tahiti) and still alive. Recognition of the above-threshold exposures of this population in the wake of the Centaure test would pave the way for this group to receive compensation under the *Loi Morin*.

Assuming an average population death rate of 0.54% and an average recognized cancers prevalence of ~0.2% between 1975 and 2020, we estimate the total number of cancers for this population to be ~10,000. Under the current law, both victims and next-of-kin of those who developed these cancers and died can apply for compensation until 12/31/2021.[80] The average compensation per recognized victim is about 70,000 euros.[81] Thus, the cost of compensation for these 10,000 cancer cases would be ~700 million euros. After the end of 2021, however, only people who are alive or the relatives of those who have died within the previous three years can be compensated.[82] For the years 2019 and 2020, approximately 700 cancers would qualify for a total cost of up to 49 million euros. Starting in 2021, the annual cost of compensation would be about 350 x 70,000 = 24.5 million euros per year. Beyond that, projections would have to take into account age-related increases in rate of cancer in a population declining due to all causes of mortality.

---

[78] Ministère de la Santé Polynésien, Données d'incidence et de prevalence des cancers, situation au 14 décembre 2017, Letter ref 2400/MSS from Jacques Raynal to Éliane Tevahitua, December 28, 2017.

[79] Ferlay J et al. (2020). Global Cancer Observatory: Cancer Today. Lyon, France: International Agency for Research on Cancer. Available from: https://gco.iarc.fr/today.

[80] LOI n° 2010-2 du 5 janvier 2010 relative à la reconnaissance et à l'indemnisation des victimes des essais nucléaires française, JORF n°0004 du 6 janvier 2010, Article 1(II).

[81] CIVEN, Rapport d'activité 2019, https://www.gouvernement.fr/sites/default/files/contenu/piece-jointe/2020/07/rapport_dactivite_2019_avec_annexes.pdf, pp. 16.

[82] LOI n° 2010-2 du 5 janvier 2010 relative à la reconnaissance et à l'indemnisation des victimes des essais nucléaires française, JORF n°0004 du 6 janvier 2010, Article 1(II).

In this study, we reviewed and reevaluated population-exposure estimates for six French atmospheric tests in light of publicly available data and recently declassified French government documents. Between 1966 and 1974, France conducted a full 41 above-ground tests. Our results are therefore far from comprehensive. They illustrate, however, some of the broader pitfalls of basing compensation decisions on dose reconstruction studies that are at best incomplete and uncertain, and at worst systematically under-representative of levels of external and internal radiation exposures.

In this context of a low dose compensation threshold coupled with high uncertainties in dose reconstructions, French policymakers should consider a pure presumption of causality applicable to all ~125,000 inhabitants present in French Polynesia at the time of the atmospheric tests (as it was briefly the case in 2017, in the immediate wake of the *Loi EROM*). Per our above analyses, given that 110,000 individuals may have been exposed to 1mSv or more of ionizing radiation in the Society Islands in 1974, this pure presumption of causality would not dramatically enlarge the pool of eligible individuals and would provide for more equitable compensation in the absence of comprehensive exposure data over all islands at all times.

# REFERENCES

List of references to be generated from footnotes.

**Data availability**: All data generated or analyzed during this study are included in the published article and are available from the corresponding author upon request.

**Acknowledgements**: The authors gratefully acknowledge the NOAA Air Resources Laboratory (ARL) for the provision of the HYSPLIT transport and dispersion model used in this publication. We thank V. Jodoin and R. Draxler for useful conversations and advices in applying Hysplit to nuclear fallout reconstructions. The authors are thankful to Z. Mian and F. von Hippel for providing valuable feedback on the manuscript. The authors also thank S. Lavrenchuk for her help with making Figures 2,3,4 and the supplementary Centaure video.

**Authors' contributions**: SP and NA conceptualized this study. SP conducted the dose analysis and the atmospheric transport modeling. SS led the legal analysis. All authors contributed to the policy analysis and writing of the manuscript.

**Competing interests**: The authors declare no competing interests.

# SUPPLEMENTARY MATERIALS for

Radiation Exposures and Compensation of Victims
from French Atmospheric Nuclear Tests in Polynesia

Sébastien Philippe[1,*] Sonya Schoenberger[2,3], Nabil Ahmed[4]

[1] Program on Science and Global Security, Princeton University, 221 Nassau St, 2nd floor, Princeton NJ 08540
[2] Department of History, Stanford University, 450 Serra Mall Bldg 200, Stanford, CA 94305
[3] Yale Law School, 127 Wall St, New Haven, CT 06511
[4] Faculty of Architecture and Design, Norwegian University of Science and Technology, Alfred Getz vei 3, 7491 Trondheim

*Corresponding author: sebastien@princeton.edu

**This PDF file includes:**

Methods
- Atmospheric transport simulation of nuclear test fallouts
- Water contamination from ALDEBARAN fallout on Gambier

List of French atmospheric nuclear tests in the Pacific
- Table S1

Historical Dose Estimates
- First publicly available estimates produced in 1997 (Table S2)
- 1970s dose estimates based on documents declassified in 2013 (Tables S3, S4, S5)

Re-evaluation of the 2006 CEA Dose Reconstruction Studies
- RIGEL on Gambier (Table S6)
- RIGEL on Tureia (Table S7)
- ACTURUS on Tureia (Table S8)
- ENCELADE on Tureia (Table S9)
- PHOEBE on Gambier (Table S10)
- CENTAURE on Tahiti (Tables S11 and S12)

Additional References

**METHODS**

**Atmospheric transport simulation of nuclear test fallout**

Fallout patterns and cloud trajectories from French nuclear tests were reconstructed using the US NOAA Hybrid Single-Particle Integrated Trajectory (HYSPLIT) particle transport and dispersion model.[83] Following previous studies,[84,85] we modeled the dispersion and deposition of fallout from a stabilized nuclear cloud using data available in historical declassified documents or French government publications and meteorological data from the NCEP/NCAR Reanalysis (1948 - present) project.[86]

Initial clouds are represented as a vertical linear source with activity distributed among the cap, skirt and stem of the cloud as 0.775/0.15/0.075 for barge surface tests and 0.9712/0.0283/0.0005 for balloon tests. The cloud particle sizes were assumed to be log-normally distributed with parameters (d=73.6μm, s=1.51) for barge tests and (d=0.15, s=2.5) for balloon tests. Particles were equally distributed in 100 diameter bins of equal activity (according to the 2.5 and 3$^{rd}$ moment respectively) and summed to a unit release. Particle density was kept constant over time and was assumed to range from 1.30–2.16 to 4.8 g/cm$^3$ for barge and balloon tests respectively. The particle density for the CENTAURE test was obtained from a granulometric measurement of particles collected at the Tahiti Mahina radiological measurement station.[87]

Hysplit calculations were run on a 12-core linux machine with MPI and involved the release of 3,000,000 3D particles each. Output data include particle air concentration between 0 and 500m and ground deposition. Air and ground activity were then computed assuming no fractionation of fission products using the con2rem Hysplit routine. The source code of con2rem was modified to allow for the use of more than 500 nuclides in the source term and to account for the beginning of decay at the time of the explosion and not at the beginning of the simulation. Different source terms were used for activity concentration maps or dose rate contour calculations normalized at H+1 using uranium-235 and plutonium-239 ENDF8 fission product yields at 500keV and 14 MeV.[88] Decay of the source term was conducted at different time interval with Onix, an open source depletion code developed at Princeton University.[89] Activity to dose ratios were obtained from the US FGR12 and ICRP Publication 72.[90,91] A python routine was written to generate relevant con2rem activity input files.

---

[83] Stein, A.F., Draxler, R.R, Rolph, G.D., Stunder, B.J.B., Cohen, M.D., and Ngan, F., (2015). NOAA's HYSPLIT atmospheric transport and dispersion modeling system, Bull. Amer. Meteor. Soc., 96, 2059-2077, http://dx.doi.org/10.1175/BAMS-D-14-00110.1this link opens in a new window

[84] Moroz, Brian E., Harold L. Beck, André Bouville, and Steven L. Simon. "Predictions of dispersion and deposition of fallout from nuclear testing using the NOAA-HYSPLIT meteorological model." Health physics 99, no. 2 (2010).

[85] Rolph, G. D., F. Ngan, and R. R. Draxler. "Modeling the fallout from stabilized nuclear clouds using the HYSPLIT atmospheric dispersion model." Journal of environmental radioactivity 136 (2014): 41-55.

[86] Kalnay et al.,The NCEP/NCAR 40-year reanalysis project, Bull. Amer. Meteor. Soc., 77, 437-470, 1996.

[87] SMSR, "Retombées consecutives au tir CENTAURE," Report 101/SMSR.PAC/CD, September 7, 1974

[88] Brown, David A., M. B. Chadwick, R. Capote, A. C. Kahler, A. Trkov, M. W. Herman, A. A. Sonzogni et al. "ENDF/B-VIII. 0: the 8th major release of the nuclear reaction data library with CIELO-project cross sections, new standards and thermal scattering data." Nuclear Data Sheets 148 (2018): 1-142.

[89] J. de Troullioud de Lanversin, M. Kütt, A. Glaser, ONIX: An open-source depletion code, Annals of Nuclear Energy, Volume 151, 2021, 107903, ISSN 0306-4549, https://doi.org/10.1016/j.anucene.2020.107903.

While fallout dose contours were found to be relatively similar to those produced historically by the French government, we found that air activity concentrations computed by Hysplit were under-predicted by a factor of 5 to 10. More detailed calculations and discussion of the challenges and opportunities of modelling lagoon and air bursts at close and long ranges up to several weeks will be the focus of a future publication.

**Water contamination from ALDEBARAN fallout on Gambier**

The contamination of rainwater from the ALDEBARAN test is estimated from two different type of measures, both obtained from historical documents.

First, the radiological survey team, measured the contamination levels of rainwater samples collected on 7/8/66 and 7/9/66 and found values of 16650 Bq/l and 14800 Bq/l respectively.[92] These correspond to values at H+11 ranging from $(14000/814)*14800 = 2.54 \cdot 10^5$ Bq/l to $(11/(7*24+11))^{-1.2} * 14800 = 4.2 \cdot 10^5$ Bq/l depending on methodologies to account for decay between H+11 and the sampling time. This gives a factor of ~20 difference between the water contamination measured in Rikitea and the contaminated rainwater.

Second, on July 3rd, activity in rainwater samples collected from the SMSR pluviometer ranged from $5.48 \cdot 10^5$ to $1.39 \cdot 10^6$ Bq/l.[93] Correcting for decay this gives $2.19 \cdot 10^6$ to $5.56 \cdot 10^6$ Bq/l at H+11. The pluviometer collection area was 1 m$^2$ and it rained 13mm. Typical individual household water collection involved surfaces of 30 m$^2$ and cisterns of 15000 liters capacity.[94] Remembering that the measured total deposited activity was $6.2 \cdot 10^7$ Bq/m$^2$, this means that to a first order about 390 liters of contaminated water ($6.2 \cdot 10^7/(0.013*1000) = 4.77 \cdot 10^6$ Bq/l at H+11) would have been collected on the first day. Assuming a cistern half full, this activity would have been diluted by a factor of ~20 (note that the cistern could have been at lesser capacity) to ~$2.5 \cdot 10^5$ Bq/l or 20 times the activity of the Rikitea water system.

---

[90] Eckerman, Keith F., and Jeffrey Clair Ryman. External exposure to radionuclides in air, water, and soil. No. CONF-960415--36. Oak Ridge National Lab., 1996.
[91] ICRP, 1995. Age-dependent Doses to the Members of the Public from Intake of Radionuclides - Part 5 Compilation of Ingestion and Inhalation Coefficients. ICRP Publication 72. Ann. ICRP 26 (1).
[92] SMCB, "Rapport préliminaire concernant les résultats obtenus par le BRO "La Coquille" pendant la 1ere demi-campagne," Report 110/CEP/SMCB/S, August 165, 1966.
[93] SMSR, "Bilan des mesures physiques concernant Mangareva," Repor 20/SMSR/PEL/PAC/S August 30, 1966.
[94] Drozdovitch, Vladimir et al. "Thyroid Doses to French Polynesians Resulting from Atmospheric Nuclear Weapons Tests: Estimates Based on Radiation Measurements and Population Lifestyle Data." *Health Physics* 120, no. 1 (2021): 34-55.

**Table S1.** List of French atmospheric nuclear tests in the Pacific (1966-1974).

| # | Date | Time | Name | Location | Type | Height [m] | Yield [kT] | Type |
|---|---|---|---|---|---|---|---|---|
| 1 | 7/2/66 | 15:34 | ALDÉBARAN | Mururoa Dindon | Barge | 10 | 28 | Fission |
| 2 | 7/19/66 | 15:05 | TAMOURÉ | 85 km E Mururoa | Air drop | 1000 | 50 | Fission |
| 3 | 7/21/66 | 12:00 | GANYMÈDE | Mururoa Colette | Tower | 12 | 0 | Fission - safety |
| 4 | 9/11/66 | 17:30 | BÉTELGEUSE | Mururoa Denise | Balloon | 470 | 110 | Fission |
| 5 | 9/24/66 | 17:00 | RIGEL | Fangataufa Frégate | Barge | 3 | 125 | Fission |
| 6 | 10/4/66 | 21:00 | SIRIUS | Mururoa Dindon | Barge | 10 | 205 | Fission |
| 7 | 6/5/67 | 19:00 | ALTAÏR | Mururoa Denise | Balloon | 295 | 15 | Fission |
| 8 | 6/27/67 | 18:30 | ANTARÈS | Mururoa Dindon | Balloon | 340 | 120 | Fission |
| 9 | 2/7/67 | 17:30 | ARCTURUS | Mururoa Denise | Barge | 3 | 22 | Fission |
| 10 | 7/7/68 | 22:00 | CAPELLA | Mururoa Denise | Balloon | 463 | 115 | Fission |
| 11 | 7/15/68 | 19:00 | CASTOR | Mururoa Dindon | Balloon | 650 | 450 | Fission |
| 12 | 8/3/68 | 21:00 | POLLUX | Mururoa Denise | Balloon | 490 | 150 | Fission |
| 13 | 8/24/68 | 18:30 | CANOPUS | Fangataufa Frégate | Balloon | 520 | 2600 | Fission+Fusion |
| 14 | 9/8/68 | 19:00 | PROCYON | Mururoa Dindon | Balloon | 700 | 1280 | Fission+Fusion |
| 15 | 5/15/70 | 18:00 | ANDROMÈDE | Mururoa Denise | Balloon | 220 | 13 | Fission |
| 16 | 5/22/70 | 18:30 | CASSIOPÉE | Mururoa Dindon | Balloon | 500 | 224 | Fission+Fusion |
| 17 | 5/30/70 | 18:00 | DRAGON | Fangataufa Frégate | Balloon | 500 | 945 | Fission+Fusion |
| 18 | 6/24/70 | 18:30 | ERIDAN | Mururoa Denise | Balloon | 220 | 12 | Fission |
| 19 | 7/3/70 | 18:30 | LICORNE | Mururoa Dindon | Balloon | 500 | 914 | Fission+Fusion |
| 20 | 7/27/70 | 19:00 | PÉGASE | Mururoa Denise | Balloon | 220 | 0.05 | Fission |
| 21 | 8/2/70 | 19:00 | ORION | Fangataufa Frégate | Balloon | 400 | 72 | Fission+Fusion |
| 22 | 8/6/70 | 19:00 | TOUCAN | Mururoa Dindon | Balloon | 500 | 594 | Fission+Fusion |
| 23 | 6/5/71 | 19:15 | DIONÉ | Mururoa Denise | Balloon | 275 | 34 | Fission |
| 24 | 6/12/71 | 19:15 | ENCELADE | Mururoa Dindon | Balloon | 450 | 440 | Fission |
| 25 | 7/4/71 | 21:30 | JAPET | Mururoa Denise | Balloon | 230 | 9 | Fission |
| 26 | 8/8/71 | 18:30 | PHOEBÉ | Mururoa Denise | Balloon | 230 | 4 | Fission |
| 27 | 8/14/71 | 19:00 | RHÉA | Mururoa Dindon | Balloon | 480 | 955 | Fission+Fusion |
| 28 | 6/25/72 | 19:00 | UMBRIEL | Mururoa Denise | Balloon | 230 | 0.5 | Fission |
| 29 | 6/30/72 | 18:30 | TITANIA | Mururoa Dindon | Balloon | 220 | 4 | Fission |
| 30 | 7/27/72 | 18:40 | OBÉRON | Mururoa Dindon | Balloon | 220 | 6 | Fission |
| 31 | 7/31/72 | 22:30 | ARIEL | Mururoa Colette | Tower | 10 | 0.001 | Fission - safety |
| 32 | 7/21/73 | 18:00 | EUTERPE | Mururoa Dindon | Balloon | 220 | 11 | Fission |
| 33 | 7/28/73 | 23:06 | MELPOMÈNE | Mururoa Denise | Balloon | 270 | 0.05 | Fission |
| 34 | 8/18/73 | 18:15 | PALLAS | Mururoa Denise | Balloon | 270 | 4 | Fission |
| 35 | 8/24/73 | 18:00 | PARTHÉNOPE | Mururoa Dindon | Balloon | 220 | 0.2 | Fission |
| 36 | 8/28/73 | 18:30 | TAMARA | 26km W Mururoa | Air drop | 250 | 6 | Fission |
| **37** | 9/13/73 | 15:42 | VESTA | Mururoa Colette | Tower | 4.1 | 0 | Fission - safety |
| **38** | 6/16/74 | 17:30 | CAPRICORNE | Mururoa Dindon | Balloon | 220 | 4 | Fission |
| **39** | 7/1/74 | 17:30 | BÉLIER | Mururoa Colette | Tower | 5.6 | 0 | Fission - safety |
| **40** | 7/7/74 | 23:15 | GÉMEAUX | Mururoa Dindon | Balloon | 312 | 150 | Fission+Fusion |
| **41** | 7/17/74 | 17:00 | CENTAURE | Mururoa Denise | Balloon | 270 | 4 | Fission |
| **42** | 7/25/74 | 17:30 | MAQUIS | 17 km O-S-O Mururoa | Air drop | 250 | 8 | Fission |
| **43** | 7/28/74 | 17:30 | PERSÉE | Mururoa Colette | Tower | 5.6 | 0.001 | Fission - safety |
| **44** | 8/14/74 | 0:30 | SCORPION | Mururoa Dindon | Balloon | 312 | 96 | Fission+Fusion |
| **45** | 8/24/74 | 23:45 | TAUREAU | Mururoa Denise | Balloon | 270 | 14 | Fission |
| **46** | 9/14/74 | 23:30 | VERSEAU | Mururoa Dindon | Balloon | 433 | 332 | Fission+Fusion |

# HISTORICAL DOSE ESTIMATES

## First publicly available estimates produced in 1997

**Table S2.** Effective doses from selected French atmospheric tests, 1997 CEA/DAM estimate (Bourges, 1997, p.940)

| UNIT = mSV | **ALD** | **ARC** | **ENC** | **PHO** | **CEN** |
|---|---|---|---|---|---|
| **LOCATION** | GAMBIER | TUREIA | TUREIA | GAMBIER | TAHITI |
| **External exposition** | 3.40 | 0.70 | 0.90 | 0.90 | 0.60 |
| **Inhalation** | 0.18 | 0.02 | 0.00 | 0.00 | 0.08 |
| **Ingestion** | 1.90 | 0.17 | 0.43 | 0.24 | 0.06 |
| **Total** | 5.48 | 0.89 | 1.33 | 1.14 | 0.74 |
| Official Rounding | **5.50** | **0.90** | **1.30** | **1.20** | **0.80** |

## 1970s dose estimates based on documents declassified in 2013

Recently declassified documents show that estimates of external and internal doses to populations were computed as early as the first atmospheric test (ALDEBARAN, July 1966). Two documents provide summary of external and internal exposures for the atmospheric tests period.[95,96] Additional documents provide estimate of internal contamination due to the presence of long-lived fission products in the food chain from 1975 to 1988. We extracted these data and combined them to obtain the official dose estimates known in the 1970s (See Table S3 and S4). By combining dose for newborns and children we also produce total dose estimates for children based on their birth year (See Table S5). All these estimates predate the first public estimates from Bourges (1997) by twenty years.

Overall, the historical data confirm the impact of atmospheric tests on the populations of Tahiti, Gambier, and Tureia. While the 2006 retrospective dose reconstructions sometimes led to higher doses, this was not always the case in particular for Papetee.

---

[95] SMSR, Compte Rendu de la Campagne 1974, Report 11/SMSR/DIR/SD, November 26, 1974, p. 91
[96] SMCB, Controle Biologique en Polynésie 1975-1976, Vol2, 1978, p. 186-190

**Table S3.** Annual effective dose (1970s estimates) to a newborn, a 7 year-old, and an adult in the islands of Tahiti (Papeete), Tureia, and Gambier for the 1966-74 atmospheric test period. The doses are obtained by summing SMSR external dose estimates and SMCB internal dose estimates. Dose from inhalation were not computed. Internal doses include the ingestion of radionuclides such as Cs-137, Sr-90 and Co-60. Dose to newborn does not include exposure in utero.

| Effective Dose | **Papeete** | | | **Tureia** | | | **Gambier** | | |
|---|---|---|---|---|---|---|---|---|---|
| [mSv] | newborn | child | adult | newborn | child | adult | newborn | child | adult |
| 1966 | 0.11 | 0.03 | 0.02 | 0.95 | 0.95 | 0.67 | 3.97 | 7.01 | 4.94 |
| 1967 | 0.05 | 0.03 | 0.02 | 1.71 | 1.82 | 1.55 | 0.14 | 0.24 | 0.17 |
| 1968 | 0.05 | 0.02 | 0.02 | 0.21 | 0.19 | 0.19 | 0.03 | 0.03 | 0.03 |
| 1969 | 0.02 | 0.02 | 0.02 | 0.10 | 0.10 | 0.10 | 0.02 | 0.02 | 0.02 |
| 1970 | 0.08 | 0.03 | 0.02 | 0.18 | 0.21 | 0.19 | 0.21 | 0.21 | 0.21 |
| 1971 | 0.15 | 0.07 | 0.06 | 2.16 | 2.16 | 1.92 | 1.88 | 2.50 | 2.08 |
| 1972 | 0.02 | 0.02 | 0.02 | 0.09 | 0.09 | 0.09 | 0.02 | 0.02 | 0.02 |
| 1973 | 0.11 | 0.05 | 0.05 | 0.05 | 0.05 | 0.05 | 0.02 | 0.02 | 0.02 |
| 1974 | 1.46 | 1.21 | 1.16 | 0.12 | 0.12 | 0.12 | 0.04 | 0.04 | 0.04 |
| TOTAL [mSv] | | 1.48 | 1.39 | | 5.69 | 4.88 | | 10.11 | 7.54 |

**Table S4.** Annual thyroid dose (1970s estimates) to a newborn, a 7 year-old child, and an adult in the islands of Tahiti (Papeete), Tureia, and Gambier for the 1966-74 atmospheric test period. The doses are obtained by summing SMSR external dose estimates and SMCB internal dose estimates. Dose from inhalation were not computed. Internal doses include the ingestion of radioiodine through contaminated water. Dose to newborn does not include exposure in utero.

| Thyroid Dose | **Papeete** | | | **Tureia** | | | **Gambier** | | |
|---|---|---|---|---|---|---|---|---|---|
| [mSv] | newborn | child | adult | newborn | child | adult | newborn | child | adult |
| 1966 | 2 | 0.26 | 0.1 | 10.1 | 10.14 | 3.9 | 7.4 | 75.3 | 29 |
| 1967 | 0.69 | 0.09 | 0.03 | 7.36 | 9.88 | 3.8 | 0.23 | 2.3 | 0.89 |
| 1968 | 0.59 | 0.09 | 0.04 | 0.59 | 0.09 | 0.03 | 0 | 0.11 | 0.04 |
| 1969 | 0 | 0 | 0 | 0 | 0 | 0 | 0 | 0 | 0 |
| 1970 | 1.34 | 0.19 | 0.07 | 0 | 0.65 | 0.25 | 0 | 0 | 0 |
| 1971 | 2.13 | 0.28 | 0.11 | 8.46 | 8.49 | 3.27 | 1.52 | 15.45 | 5.94 |
| 1972 | 0.12 | 0.02 | 0.01 | 0 | 0 | 0 | 0 | 0 | 0 |
| 1973 | 1.42 | 0.2 | 0.08 | 0 | 0 | 0 | 0 | 0 | 0 |
| 1974 | 7.44 | 2.04 | 0.78 | 0 | 0 | 0 | 0 | 0 | 0 |
| TOTAL [mSv] | | 3.17 | 1.22 | | 29.25 | 11.25 | | 93.16 | 35.87 |

**Table S5.** Total effective and thyroid doses (1970s estimates) to children born and raised in the islands of Tahiti (Papeete), Tureia, and Gambier during the 1966-1974 atmospheric test period. The doses are obtained by summing the Table S12 data per birth year. Dose from inhalation were not computed. Doses do not include exposure in utero.

| Birth year | Total Effective Dose 1966-74 | | | Total Thyroid Dose 1966-74 | | | Total Effective Dose 1966-88 | | |
|---|---|---|---|---|---|---|---|---|---|
| | Papeete | Tureia | Gambier | Papeete | Tureia | Gambier | Papeete | Tureia | Gambier |
| 1965 | 1.48 | 5.69 | 10.11 | 3.17 | 29.25 | 93.16 | 1.54 | 5.89 | 10.19 |
| 1966 | 2.29 | 5.69 | 7.07 | 4.91 | 29.21 | 25.26 | 2.36 | 5.89 | 7.15 |
| 1967 | 1.80 | 4.62 | 3.00 | 3.51 | 16.59 | 15.79 | 1.87 | 4.82 | 3.08 |
| 1968 | 1.67 | 2.94 | 2.85 | 3.32 | 9.73 | 15.45 | 1.74 | 3.14 | 2.93 |
| 1969 | 1.47 | 2.72 | 2.82 | 2.73 | 9.14 | 15.45 | 1.53 | 2.92 | 2.90 |
| 1970 | 1.43 | 2.60 | 2.80 | 3.88 | 8.49 | 15.45 | 1.50 | 2.80 | 2.88 |
| 1971 | 1.74 | 2.42 | 1.96 | 4.39 | 8.46 | 1.52 | 1.81 | 2.62 | 2.04 |
| 1972 | 1.32 | 0.26 | 0.08 | 2.36 | 0.00 | 0.00 | 1.38 | 0.46 | 0.16 |
| 1973 | 1.32 | 0.17 | 0.07 | 3.46 | 0.00 | 0.00 | 1.39 | 0.37 | 0.15 |
| 1974 | 1.46 | 0.12 | 0.04 | 7.44 | 0.00 | 0.00 | 1.52 | 0.32 | 0.12 |
| 1975 | | | | | | | 0.07 | 0.20 | 0.08 |

# RE-EVALUATION OF THE 2006 CEA DOSE RECONSTRUCTIONS

RIGEL on GAMBIER (7/2/1966)

**Table S6.** Effective and thyroid dose estimates for the RIGEL fallout on the Gambier archipelago. Results are based on 2006 CEA data and our re-evaluations. Updated values in bold include corrections to the groundshine dose as well as doses from contaminated water consumption. Groundshine and cloudshine contributions to the thyroid, originally missing, were also included.

|  | EFFECTIVE DOSE [mSv] | | | | THYROID DOSE [mSv] | | | |
|---|---|---|---|---|---|---|---|---|
|  | Child age 1-2 | | Adult | | Child age 1-2 | | Adult | |
|  | Min | Max | Min | Max | Min | Max | Min | Max |
| Groundshine | **0.02** | **0.05** | **0.02** | **0.05** | **0.02** | **0.04** | **0.02** | **0.04** |
| Cloudshine | 0.00 | **0.00** | 0.00 | **0.00** | 0.00 | **0.00** | 0.00 | **0.00** |
| Inhalation | 0.00 | **0.00** | 0.00 | **0.00** | 0.01 | **0.02** | 0.01 | **0.01** |
|  |  |  |  |  |  |  |  |  |
| Water | **0.38** | **7.91** | **0.10** | **2.10** | **4.40** | **90.21** | **1.00** | **21.01** |
| Vegetables | 0.00 | **0.05** | 0.00 | **0.04** | 0.02 | **0.59** | 0.01 | **10.93** |
| Fish and Mollusc | 0.01 | **0.02** | 0.01 | **0.02** | 0.13 | **0.20** | 0.10 | **0.15** |
|  |  |  |  |  |  |  |  |  |
| Ingestion | 0.39 | 7.98 | 0.11 | 2.16 | 4.55 | 91.00 | 1.11 | 32.10 |
|  |  |  |  |  |  |  |  |  |
| Internal | 0.40 | 7.98 | 0.11 | 2.16 | 4.56 | 91.02 | 1.11 | 32.10 |
| External | 0.02 | 0.05 | 0.02 | 0.05 | 0.02 | 0.04 | 0.02 | 0.04 |
|  |  |  |  |  |  |  |  |  |
| **Total [mSv]** | **0.42** | **8.03** | **0.13** | **2.20** | **4.58** | **91.06** | **1.13** | **32.15** |
| CEA 2006 | 0.41 | 0.71 | 0.13 | 0.23 | 4.61 | 7.81 | 1.11 | 2.11 |
| Ratio | 1.00 | 11.37 | 1.01 | 9.68 | 0.99 | 11.66 | 1.02 | 15.27 |

Corrections: The maximum groundshine dose was computed for 1 year instead of 6 months (factor of 1.04) and assumed 100% of time spent outside (factor of 1.5). Groundshine was also corrected to account for the availability of higher rain activity measurement value (5000 pCi/cm$^3$) leading to an increase by a factor of 1.52. The rain activity correction was also applied to cloudshine, inhalation as well as ingestion contributions from fish and vegetables consumption as they were all computed from this value. The contribution of water consumption to the maximum estimate was also corrected by a factor of 7.4*1.67=12.3 to account for both higher measured rain activity and the possibility of rain (cistern) water consumption on the Gambier Archipelago (based on Arcturus data corrected for possibly higher family cisterns activity). Not that technically, the maximum value could be increased further if no contaminated rain dilution is assumed.

RIGEL on TUREIA (7/2/1966)

Table S7. Effective and thyroid dose estimates for the RIGEL fallout on the Tureia atoll. Results are based on 2006 CEA data and our re-evaluations. Updated values in bold include corrections to the groundshine dose as well as doses from contaminated water consumption. Groundshine and cloudshine contributions to the thyroid, originally missing, are also included.

|  | EFFECTIVE DOSE [mSv] | | | | THYROID DOSE [mSv] | | | |
|---|---|---|---|---|---|---|---|---|
|  | Child age 1-2 | | Adult | | Child age 1-2 | | Adult | |
|  | Min | Max | Min | Max | Min | Max | Min | Max |
| Groundshine | **0.05** | **0.22** | **0.05** | **0.22** | **0.05** | **0.21** | **0.05** | **0.21** |
| Cloudshine | 0.00 | **0.00** | 0.00 | **0.00** | **0.00** | **0.00** | **0.00** | **0.00** |
| Inhalation | 0.00 | **0.01** | 0.00 | **0.01** | 0.03 | **0.09** | 0.02 | **0.04** |
|  |  |  |  |  |  |  |  |  |
| Water | 0.05 | **3.37** | 0.03 | **0.84** | 0.52 | **37.12** | 0.12 | **8.44** |
| Fish and Mollusc | 0.00 | **0.27** | 0.00 | **0.21** | 0.06 | **3.14** | 0.02 | **2.08** |
|  |  |  |  |  |  |  |  |  |
| Ingestion | 0.05 | 3.65 | 0.03 | 1.05 | 0.58 | 40.25 | 0.14 | 10.52 |
|  |  |  |  |  |  |  |  |  |
| Internal | 0.06 | 3.65 | 0.03 | 1.06 | 0.61 | 40.34 | 0.15 | 10.56 |
| External | 0.05 | 0.22 | 0.05 | 0.22 | 0.05 | 0.21 | 0.05 | 0.21 |
|  |  |  |  |  |  |  |  |  |
| **Total [mSv]** | **0.11** | **3.88** | **0.08** | **1.28** | **0.66** | **40.55** | **0.20** | **10.77** |
| CEA 2006 | 0.11 | 0.23 | 0.06 | 0.15 | 0.61 | 2.01 | 0.15 | 0.95 |
| Ratio | 1.02 | 17.02 | 1.32 | 8.78 | 1.08 | 20.17 | 1.32 | 11.40 |

Corrections: The maximum groundshine dose was computed for 1 year instead of 6 months (factor of 1.04) and assumed 100% of time spent outside (factor of 1.5). Groundshine was also corrected to account for the availability of a higher rain activity measurement value (2000 pCi/cm$^3$ measured on 9/26/1966) leading to an increase by a factor of 2.85. The rain activity correction was also applied to cloudshine, inhalation as well as foodstuffs consumption as they were all computed from this value. The contribution of water consumption to the maximum estimate was also corrected by a factor of 7.4*2.85=21.09 to account for higher measured rain activity and lower dilution of rainwater activity in family cisterns (based on Arcturus cistern data). Not that technically, the maximum value could be increased further if no dilution took place. Finally, we correct water consumption for Tureia by a factor of 2 to be coherent with recent dose reconstruction studies (see main article).

ARCTURUS on TUREIA (1967)

**Table S8.** Effective and thyroid dose estimates for the ARCTURUS fallout on the Tureia atoll. Results are based on 2006 CEA data and our re-evaluations. Updated values in bold include corrections to the groundshine dose as well as doses from contaminated water consumption. Groundshine and cloudshine contributions to the thyroid, originally missing, are also included.

|  | EFFECTIVE DOSE [mSv] | | | | THYROID DOSE [mSv] | | | |
|---|---|---|---|---|---|---|---|---|
|  | Child age 1-2 | | Adult | | Child age 1-2 | | Adult | |
|  | Min | Max | Min | Max | Min | Max | Min | Max |
| Groundshine | **0.73** | **3.28** | **0.73** | **3.28** | **0.69** | **3.11** | **0.69** | **3.11** |
| Cloudshine | 0.00 | 0.01 | 0.00 | 0.01 | **0.00** | **0.01** | **0.00** | **0.01** |
| Inhalation | 0.02 | 0.11 | 0.01 | 0.07 | 0.23 | 1.38 | 0.10 | 0.63 |
|  | | | | | | | | |
| Cistern (rain) water | 0.11 | **1.63** | 0.03 | **0.44** | 1.24 | **18.35** | 0.29 | **4.29** |
| Fruits | 0.02 | 0.02 | 0.01 | 0.01 | 0.20 | 0.22 | 0.13 | 0.14 |
| Fish | 0.02 | 0.25 | 0.01 | 0.10 | 0.22 | 2.80 | 0.08 | 1.00 |
| Mollusc | 0.03 | 2.80 | 0.03 | 2.28 | 0.34 | 31.80 | 0.26 | 22.50 |
|  | | | | | | | | |
| Ingestion | 0.18 | 4.70 | 0.08 | 2.84 | 2.00 | 53.17 | 0.76 | 27.93 |
|  | | | | | | | | |
| Internal | 0.20 | 4.81 | 0.09 | 2.91 | 2.23 | 54.55 | 0.86 | 28.56 |
| External | 0.73 | 3.29 | 0.73 | 3.29 | 0.69 | 3.12 | 0.69 | 3.12 |
|  | | | | | | | | |
| **Total [mSv]** | **0.93** | **8.09** | **0.82** | **6.19** | **2.92** | **57.67** | **1.55** | **31.68** |
| CEA 2006 | 0.90 | 4.00 | 0.79 | 3.20 | 2.23 | 37.44 | 0.86 | 24.56 |
| Ratio | 1.03 | 2.02 | 1.04 | 1.93 | 1.31 | 1.54 | 1.81 | 1.29 |

Corrections: The maximum groundshine dose was computed for 1 year instead of 6 months (factor of 1.04) and assumed 100% of time spent outside (factor of 1.5). Groundshine was also corrected to account for the availability of a measured dose rate (3mrad/h measured on 7/2/1967), leading to an increase by a factor of 3 (note that another instrument measured 5mrad/h). The rain activity correction was also applied to cloudshine, inhalation as well as foodstuffs consumption as they were all computed from this value. The contribution of water consumption to the maximum estimate was also corrected by a factor of 7.4 to account for potentially lower dilution of rainwater activity in family cisterns (as measured after the Encelade test). Not that technically, the maximum value could be increased further if no dilution took place. Finally, we correct water consumption for Tureia by a factor of 2 to be coherent with recent dose reconstruction studies (see main article).

ENCELADE on TUREIA (1971)

**Table S9.** Effective and thyroid dose estimates for the ENCELADE fallout on the Tureia atoll. Results are based on 2006 CEA data and our re-evaluations. Updated values in bold include corrections to the groundshine dose as well as doses from contaminated water and mollusc consumption. Groundshine and cloudshine contributions to the thyroid, originally missing, were also included.

|  | EFFECTIVE DOSE [mSv] | | | | THYROID DOSE [mSv] | | | |
|---|---|---|---|---|---|---|---|---|
|  | Child age 1-2 | | Adult | | Child age 1-2 | | Adult | |
|  | Min | Max | Min | Max | Min | Max | Min | Max |
| Groundshine | **1.39** | **6.93** | **1.39** | **6.93** | **1.30** | **6.51** | **1.30** | **6.51** |
| Cloudshine | 0.00 | **0.02** | 0.00 | **0.02** | 0.00 | 0.02 | 0.00 | 0.02 |
| Inhalation | 0.01 | **0.12** | 0.00 | **0.05** | 0.14 | **1.64** | **0.04** | **0.50** |
|  |  |  |  |  |  |  |  |  |
| Cistern (rain) water | 0.25 | **3.82** | 0.06 | **0.95** | 3.00 | **44.73** | 0.66 | **9.96** |
| Coconut water | 0.01 | 0.01 | 0.00 | 0.00 | 0.02 | 0.02 | 0.01 | 0.01 |
| Vegetables (papaye) | 0.00 | 0.00 | 0.00 | 0.00 | 0.01 | 0.01 | 0.01 | 0.01 |
| Vegetables (coprah) | 0.01 | 0.01 | 0.00 | 0.00 | 0.03 | 0.03 | 0.01 | 0.01 |
| Meat | 0.00 | 0.01 | 0.00 | 0.00 | 0.01 | 0.06 | 0.00 | 0.02 |
| Fish | 0.01 | 0.24 | 0.01 | 0.10 | 0.13 | 2.25 | 0.05 | 0.84 |
| Mollusc | 0.07 | **0.38** | 0.06 | **0.32** | 0.54 | **3.56** | 0.41 | **2.69** |
|  |  |  |  |  |  |  |  |  |
| Ingestion | 0.35 | 4.46 | 0.13 | 1.38 | 3.74 | 50.66 | 1.14 | 13.53 |
|  |  |  |  |  |  |  |  |  |
| Internal | 0.36 | 4.59 | 0.13 | 1.43 | 3.88 | 52.30 | 1.18 | 14.03 |
| External | 1.39 | 6.94 | 1.39 | 6.94 | 1.30 | 6.53 | 1.30 | 6.53 |
|  |  |  |  |  |  |  |  |  |
| **Total [mSv]** | **1.75** | **11.53** | **1.52** | **8.37** | **5.18** | **58.82** | **2.49** | **20.56** |
| CEA 2006 | 1.49 | 3.50 | 1.25 | 1.91 | 3.88 | 26.54 | 1.18 | 7.53 |
| Ratio | 1.18 | 3.30 | 1.21 | 4.38 | 1.34 | 2.22 | 2.10 | 2.73 |

Corrections: The maximum groundshine dose was computed for 1 year instead of 6 months (factor of 1.04). It assumed 100% of time spent outside (factor of 1.5) and took into account the first six hours of the fallout (+0.32 mSv). It was also corrected to account for the difference between the reconstructed and the measured dose rate (6mrad/h measured on 6/13/1971), leading to an increase by a factor of 3.33. Maximum inhalation and cloudshine were multiplied by 2 (as the CEA assumed a 0.5 factor because the fallout occurred at night). Internal dose from mollusc consumption was multiplied by 1.58 to account for the maximum activity measured in Benitier (measurement ref. 38628A). The maximum dose from water was multiplied by 1.06 to account for the maximum measured value. Finally, we corrected water consumption for Tureia by a factor of 2 to be coherent with recent dose reconstruction studies (see main article).

PHOEBE on GAMBIER (1971)

**Table S10.** Effective and thyroid dose estimates for the PHOEBE fallout on the Gambier archipelago. Results are based on 2006 CEA data and our re-evaluations. Updated values in bold include corrections to the groundshine dose as well as doses from contaminated water consumption. Groundshine and cloudshine contributions to the thyroid, originally missing, were also included.

|  | EFFECTIVE DOSE [mSv] | | | | THYROID DOSE [mSv] | | | |
|---|---|---|---|---|---|---|---|---|
|  | Child age 1-2 | | Adult | | Child age 1-2 | | Adult | |
|  | Min | Max | Min | Max | Min | Max | Min | Max |
| Groundshine | **0.11** | **0.65** | **0.11** | **0.65** | **0.11** | **0.62** | **0.11** | **0.62** |
| Cloudshine | 0.00 | 0.00 | 0.00 | 0.00 | **0.00** | **0.00** | **0.00** | **0.00** |
| Inhalation | 0.00 | 0.00 | 0.00 | 0.00 | 0.01 | **0.04** | 0.00 | **0.02** |
|  | | | | | | | | |
| Cistern (rain) water | 0.37 | **10.99** | 0.10 | **2.83** | 4.30 | **138.47** | 1.00 | **30.46** |
| Fruits | 0.03 | 0.72 | 0.03 | 0.64 | 0.34 | 9.00 | 0.27 | 7.10 |
| Fish | 0.00 | 0.00 | 0.00 | 0.00 | 0.00 | 0.00 | 0.00 | 0.00 |
| Mollusc | 0.02 | 0.05 | 0.01 | 0.02 | 0.18 | 0.58 | 0.06 | 0.20 |
|  | | | | | | | | |
| Ingestion | 0.41 | 11.76 | 0.13 | 3.49 | 4.82 | 148.05 | 1.33 | 37.76 |
|  | | | | | | | | |
| Internal | 0.41 | 11.76 | 0.13 | 3.49 | 4.83 | 148.09 | 1.33 | 37.78 |
| External | 0.11 | 0.65 | 0.12 | 0.65 | 0.11 | 0.62 | 0.11 | 0.62 |
|  | | | | | | | | |
| **Total [mSv]** | **0.53** | **12.42** | **0.25** | **4.14** | **4.94** | **148.71** | **1.44** | **38.40** |
| CEA 2006 | 0.52 | 7.88 | 0.24 | 2.57 | 4.83 | 97.82 | 1.33 | 26.72 |
| Ratio | 1.01 | 1.57 | 1.02 | 1.61 | 1.02 | 1.52 | 1.08 | 1.44 |

Corrections: The maximum groundshine dose was computed for 1 year instead of 6 months (factor of 1.04). It assumed 100% of time spent outside (factor of 1.5). It was also corrected to account for the difference between the reconstructed and the measured dose rate (5mrad/h measured on 8/8/1971), leading to an increase by a factor of 3.8. The maximum dose from water consumption was multiplied by 1.57 to account for the maximum measured activity in Taku (1800 Bq of I-131 per liter of water).

CENTAURE on Tahiti (1974)

**Table S11.** Corrected effective dose estimates for the CENTAURE fallout. Updated values in bold reflect higher ground deposition for the reference station (MAHINA), higher upper bound for deposition in the Pirae/Papeete zone, absence of data from the PAPEETE/PIRAE market for vegetables. Groundshine was corrected for 1 year (instead of 6 months) and we assumed time spent outdoor to be 2/3 for Teahupoo/Taravao (as is assumed by the CEA for Pirae and Hitiaa).

| Effective Dose | Child age 1-2 | | | Adult | | |
|---|---|---|---|---|---|---|
| | Pirae | Hitiaa | Taravao | Pirae | Hitiaa | Taravao |
| Groundshine | **0.17** | **1.70** | **6.22** | **0.17** | **1.70** | **6.22** |
| Cloudshine | **0.01** | **0.03** | **0.12** | **0.01** | **0.03** | **0.12** |
| Inhalation | **0.16** | **0.78** | **2.86** | **0.15** | **0.71** | **2.58** |
| | | | | | | |
| Water | 0.05 | 0.10 | 0.02 | 0.02 | 0.03 | 0.00 |
| Milk | 0.36 | 2.10 | 0.36 | 0.03 | 0.20 | 0.03 |
| Vegetables | **0.95** | 0.95 | 0.56 | **0.41** | 0.41 | 0.24 |
| Meat | 0.08 | 0.04 | 0.04 | 0.04 | 0.02 | 0.02 |
| Eggs | 0.02 | 0.01 | 0.01 | 0.01 | 0.00 | 0.00 |
| Fish | 0.15 | 0.15 | 0.15 | 0.07 | 0.09 | 0.09 |
| Mollusc | 0.02 | 0.07 | 0.07 | 0.01 | 0.06 | 0.06 |
| Shellfish | 0.06 | 0.06 | 0.06 | 0.03 | 0.03 | 0.03 |
| | | | | | | |
| Ingestion | 1.69 | 3.48 | 1.26 | 0.61 | 0.84 | 0.47 |
| | | | | | | |
| Internal | 1.85 | 4.25 | 4.11 | 0.76 | 1.55 | 3.06 |
| External | 0.17 | 1.73 | 6.35 | 0.17 | 1.73 | 6.35 |
| | | | | | | |
| **Total [mSv]** | **2.02** | **5.98** | **10.46** | **0.93** | **3.28** | **9.40** |
| CEA 2006 | 1.19 | 5.27 | 4.55 | 0.45 | 2.59 | 3.56 |
| Ratio | 1.70 | 1.14 | 2.30 | 2.07 | 1.27 | 2.64 |

**Table S12.** Corrected thyroid dose estimates for the CENTAURE fallout on TAHITI. Updated values in bold include groundshine and cloudshine contributions to the Thyroid dose, and address the absence of data for vegetables from the PAPEETE/PIRAE market.

| Thyroid Dose | Child age 1-2 | | | Adult | | |
|---|---|---|---|---|---|---|
| | Pirae | Hitiaa | Taravao | Pirae | Hitiaa | Taravao |
| Groundshine | **0.16** | **1.61** | **5.91** | **0.16** | **1.61** | **5.91** |
| Cloudshine | **0.01** | **0.04** | **0.13** | **0.01** | **0.04** | **0.13** |
| Inhalation | **1.82** | **8.70** | **32.91** | **0.86** | **4.22** | **14.96** |
| | | | | | | |
| Water | 0.60 | 1.30 | 0.22 | 0.20 | 0.30 | 0.05 |
| Milk | 4.50 | 25.00 | 4.50 | 0.40 | 1.70 | 0.40 |
| Vegetables | **11.90** | 11.90 | 6.90 | **4.50** | 4.50 | 2.60 |
| Meat | 1.10 | 0.54 | 0.54 | 0.50 | 0.20 | 0.20 |
| Eggs | 0.18 | 0.14 | 0.07 | 0.07 | 0.05 | 0.03 |
| Fish | 1.90 | 1.90 | 1.90 | 0.80 | 1.00 | 1.00 |
| Mollusc | 0.30 | 0.84 | 0.84 | 0.12 | 0.66 | 0.66 |
| Shellfish | 0.66 | 0.66 | 0.66 | 0.30 | 0.30 | 0.30 |
| | | | | | | |
| Ingestion | 21.14 | 42.28 | 15.63 | 6.89 | 8.71 | 5.24 |
| | | | | | | |
| Internal | 22.96 | 50.98 | 48.54 | 7.75 | 12.93 | 20.20 |
| External | 0.16 | 1.65 | 6.04 | 0.16 | 1.65 | 6.04 |
| | | | | | | |
| **Total [mSv]** | **23.13** | **52.63** | **54.58** | **7.92** | **14.57** | **26.24** |
| CEA 2006 | 14.11 | 48.68 | 39.83 | 4.36 | 11.81 | 16.24 |
| Ratio | 1.64 | 1.08 | 1.37 | 1.82 | 1.23 | 1.62 |